\documentclass[%
 aps,
 physrev,
 amsmath,amssymb,
 reprint,%
superscriptaddress,%
]{revtex4-2}

\usepackage{braket}
\usepackage{graphicx}
\usepackage{dcolumn}
\usepackage{bm}

\usepackage[utf8]{inputenc}
\usepackage[T1]{fontenc}
\usepackage{mathptmx}
\usepackage{etoolbox}
\usepackage{microtype}
\usepackage{todonotes}
\setuptodonotes{noinline, size=\scriptsize}
\usepackage{siunitx}
\DeclareSIUnit{\gauss}{G}

\usepackage{afterpage}
\usepackage[version=4]{mhchem}

\let\Re\relax

\DeclareMathOperator{\Re}{Re}

\DeclareMathOperator{\tem}{TEM}
\DeclareMathAlphabet{\altmathcal}{OMS}{cmsy}{m}{n}
\newcommand{\rms}{\mathrm{rms}}

\newcommand{\mb}{\mathbf}
\newcommand{\bs}{\boldsymbol}
\newcommand{\abs}[1]{\lvert #1 \rvert}
\newcommand{\Abs}[1]{\left\lvert #1 \right\rvert}
\newcommand{\inv}{^{-1}}
\newcommand{\pa}{\partial}

\newcommand{\zr}{\ensuremath{z_R}}
\newcommand{\london}{\ensuremath{\lambda}}
\newcommand{\RRR}{\ensuremath{\mathrm{RRR}}}
\DeclareMathOperator{\tr}{tr}

\newcommand{\maxfinesse}{\num{5.8(1)e+7}}
\newcommand{\cavityNA}{\num{0.56}}
\newcommand{\etaest}{\num{2.7e+6}}

\newcommand{\mean}[1]{\overline{#1}}

\newcommand{\nearcfcl}{\ensuremath{\mean{g}_0}}

\newcommand{\lesscfcl}{\ensuremath{\mean{g}_1}}
\newcommand{\hadriana}{\ensuremath{\mean{g}_2}}
\newcommand{\hadrianalen}{\ensuremath{L_2}}
\newcommand{\nearcfclg}{\num{-0.029}}

\newcommand{\lesscfclgnum}{\num{-0.068}}
\newcommand{\hadrianagnum}{\num{-0.109}}
\newcommand{\Fmax}{F_\mathrm{max}}
\newcommand{\Fid}{\ensuremath{\altmathcal{F}}}
\newcommand{\DeltaOpt}{\ensuremath{\Delta_\mathrm{opt}}}
\newcommand{\Tgate}{\ensuremath{\tau_\mathrm{g}}}
\newcommand{\br}[1]{\ensuremath{\overline{#1}}}
\newcommand{\TC}{\mathrm{TC}}
\newcommand{\Htc}{\ensuremath{H_{\TC}}}
\newcommand{\Heff}{\ensuremath{H_\mathrm{eff}}}
\newcommand{\ketbra}[2]{\ensuremath{\lvert{#1}\rangle\langle{#2}\rvert}}

\newcommand{\geminacolor}{red}
\newcommand{\hadrianacolor}{blue}
\newcommand{\BCSNumericColor}{green}
\newcommand{\BCSBasicColor}{orange}

\usepackage{xcolor}

\usepackage[inline]{enumitem}
\usepackage{lipsum}
\usepackage{hyperref}

\usepackage{cleveref}

\usepackage[shortcuts]{extdash}

\newlist{captionenum}{enumerate*}{2}
\setlist[captionenum,1]{label=\textbf{(\alph*)}, ref={(\alph*)}}
\setlist[captionenum,2]{label=\textbf{(\roman*)}, ref={(\roman*)}}

\makeatletter
\def\@email#1#2{%
 \endgroup
 \patchcmd{\titleblock@produce}
  {\frontmatter@RRAPformat}
  {\frontmatter@RRAPformat{\produce@RRAP{*#1\href{mailto:#2}{#2}}}\frontmatter@RRAPformat}
  {}{}
}%
\makeatother
\newcommand\papertitle{Optically accessible high-finesse millimeter-wave resonator for cavity quantum electrodynamics with atom arrays}

\newcommand{\supplemental}{
  \widetext
  \begin{center}
    {}  
    \textsf{\Large{Supplemental material for: ``\papertitle{}''}}
  \end{center}
    \setcounter{section}{0}
    \renewcommand{\thesection}{S\arabic{section}}
    \setcounter{table}{0}
    \renewcommand{\thetable}{S\arabic{table}}%
    \setcounter{figure}{0}
    \renewcommand{\thefigure}{S\arabic{figure}}%
    \setcounter{equation}{0}
    \renewcommand{\theequation}{S\arabic{equation}}%
    \setcounter{page}{1}%
    \thispagestyle{empty}%
}

\newcommand\stanfordphys{Department of Physics, Stanford University, 382 Via Pueblo Mall, Stanford, California 94305, USA}
\newcommand\stanfordAP{Department of Applied Physics, Stanford University, 348 Via Pueblo Mall, Stanford, California 94305 USA}
\newcommand\SLAC{SLAC National Accelerator Laboratory, 2575 Sand Hill Road, Menlo Park, California 94025, USA}

\begin{document}

\preprint{AIP/123-QED}

\title{\papertitle{}}
\author{Tony Zhang}
\affiliation{\stanfordphys{}}%
\affiliation{\SLAC{}}%
\author{Michelle Wu}%
\author{Sam~R.\ Cohen}%
\affiliation{\stanfordphys{}}%
\author{Lin Xin}%
\affiliation{\stanfordphys{}}%
\affiliation{\SLAC{}}%
\author{Debadri Das}%
\affiliation{\SLAC{}}
\affiliation{\stanfordAP{}}
\author{Kevin~K.\,S.\ Multani}%
\affiliation{\stanfordphys{}}%
\affiliation{\SLAC{}}%
\author{Nolan Peard}%
\affiliation{\stanfordAP{}}%
\author{Anne-Marie Valente-Feliciano}%
\affiliation{
Thomas Jefferson National Accelerator Facility,
12000 Jefferson Ave, Newport News, Virginia 23606, USA
}%
\author{Paul~B.\ Welander}%
\affiliation{\SLAC{}}%
\author{Amir~H.\ Safavi-Naeini}%
\affiliation{\stanfordAP{}}%
\author{Emilio~A.\ Nanni}%
\affiliation{\SLAC{}}%
\author{Monika Schleier-Smith}
\email{schleier@stanford.edu}
\affiliation{\stanfordphys{}}%
\affiliation{\SLAC{}}%

\date{June 5, 2025}

\begin{abstract}
Cavity quantum electrodynamics (QED)
is a powerful tool in quantum science,
enabling preparation of non-classical states of light
and scalable entanglement of many atoms coupled to a single field mode.
While the most coherent atom--photon interactions
have been achieved using superconducting millimeter-wave cavities
coupled to Rydberg atoms,
these platforms so far lack the optical access 
required for trapping and addressing individual atomic qubits.
We present a millimeter-wave Fabry--P\'erot cavity with finesse~\maxfinesse{} at a temperature of \SI{1}{\K}
providing generous transverse optical access (numerical aperture~\cavityNA{}).
Conflicting goals of strong atom--photon coupling and optical access
motivate a near-confocal geometry.
Close to confocality, however, post-paraxial corrections to the cavity spectrum
introduce unexpected degeneracies between transverse modes,
leading to excess cavity loss.
Modeling these corrections
allows for tuning the cavity geometry to evade this loss,
producing a high finesse
that will enable cavity QED experiments with trapped atoms
deep in the strong coupling regime.
\end{abstract}

\maketitle


Coherent exchange of quantum information between photons and matter qubits is a key capability in quantum science,
with applications ranging from quantum sensing and networking to computation and simulation.
Achieving coherent light--matter interactions
requires placing the emitters in a cavity or waveguide
that enhances their interaction with an electromagnetic mode.
One may then use the emitters to engineer non-classical states of light~\cite{rempe1990observation,thomas2024fusion},
or use the cavity field to mediate nonlocal interactions~\cite{periwal2021programmable,helson2023density,finger2024spin,marsh2024entanglement} and entanglement~\cite{leroux2010implementation,hosten2016quantum,greve2022entanglement,cooper2024graph,barontini2015deterministic,colombo2022time,welte2018photon,dhordjevic2021entanglement} between emitters.
In ensembles of trapped atoms, the collective coupling to a single field mode has allowed for engineering squeezed~\cite{leroux2010implementation,hosten2016quantum,greve2022entanglement,cooper2024graph} and non-Gaussian~\cite{barontini2015deterministic,colombo2022time} collective spin states.
Combining strong atom--light coupling with local control in tweezer-based atom arrays~\cite{welte2018photon,dhordjevic2021entanglement,grinkemeyer2025error,deist2022mid,hu2024site} 
has further enabled pairwise entanglement of atomic qubits~\cite{welte2018photon,dhordjevic2021entanglement} and non-destructive readout for error correction~\cite{deist2022mid,hu2024site,grinkemeyer2025error}.


The coherence of atom--light interactions
is quantified by the cooperativity~$\eta = 4g^2 / \kappa\Gamma$,
which compares the vacuum Rabi frequency~$g$
to the resonator and emitter decay rates $\kappa$~and~$\Gamma$.
Achieving coherent photon-mediated interactions
between two atoms
requires operating deep in the strong coupling regime~$\eta \gg 1$.
Error rates in two-qubit gates based on cavity-mediated interactions, for example, scale as $\eta^{-1/2}$~\cite{sorensen2003measurement,jandura2024nonlocal}.
In the optical domain, the highest single-atom cooperativities achieved to date
are on the order of $\eta \sim \num{e2}$~\cite{gehr2010cavity},
posing a limit to deterministically generating high-fidelity entanglement.
Thus, the preparation of entangled states in optical cavities has required
using either atomic ensembles with collectively enhanced coupling to light~\cite{leroux2010implementation,hosten2016quantum,greve2022entanglement,cooper2024graph,barontini2015deterministic,colombo2022time}
or conditional or heralded techniques~\cite{casabone2013heralded,welte2018photon,dhordjevic2021entanglement,grinkemeyer2025error}.


Significantly higher cooperativities are possible in millimeter-wave (mm-wave) cavities, which couple to transitions between Rydberg states.
In particular, frequencies from \SIrange{50}{100}{\GHz} are low enough to couple to long-lived transitions, yet sufficiently high to suppress thermal population of cavity modes.
In this regime, a finesse~$F \gtrsim \num{e9}$ has been achieved~\cite{kuhr2007ultrahigh}
thanks to the availability of superconducting mirrors
and to mm-scale wavelengths,
long when compared to mirror surface roughness.
The resulting cooperativity~$\eta = \num{5e8}$
enabled pioneering observations of
photon-mediated interactions between atoms transiting a cavity~\cite{hagley1997generation}
and preparation of non-classical states of the cavity field~\cite{deleglise2008reconstruction} via quantum non-demolition measurements~\cite{gleyzes2007quantum}.
Microwave and mm-wave frequencies are also naturally suited for interfacing atomic and solid-state qubits~\cite{pritchard2014hybrid},
motivating demonstrations of coherent coupling between atoms and coplanar waveguide resonators~\cite{hogan2012driving,morgan2020coupling,kaiser2022cavity}.
Simultaneously coupling mm-wave and optical cavities to an atomic ensemble has furthermore enabled efficient quantum frequency conversion~\cite{suleymanzade2020tunable,kumar2023quantum}.
However, incorporating single-atom trapping and addressing into such systems remains an outstanding challenge.

\begin{figure}[t]
  \includegraphics[width=\columnwidth]{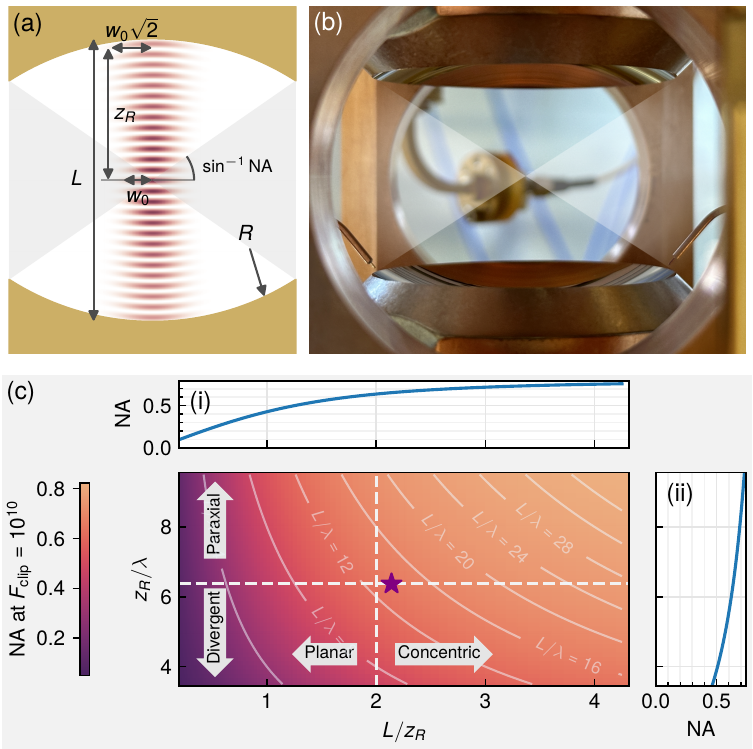}
  \caption[]{Cavity design for combining strong coupling with optical access.
      \begin{captionenum}
          \item\label{it:schematic}%
          Schematic of Fabry--P\'erot cavity of length~$L$
          and mirror radius of curvature~$R$.
          Red shading shows Gaussian $\tem_{00}$ mode
          of waist~$w_0$ and Rayleigh range~$z_R$.
          \item
          \label{it:cavity-pic}%
          The cavity and coaxial probes.
          Overlaid cones show numerical aperture (NA)
          available for imaging
          from the open sides of the cavity.
          \item\label{it:design-space}%
          Dependence of NA on dimensionless parameters~%
          $L/\zr$, describing the cavity geometry,
          and $\zr/\lambda$, characterizing the mode divergence.
          Contour lines give cavity length normalized to mode wavelength.
          Color shows NA when mirrors are sized to permit a finesse~$F_\text{clip} = 10^{10}$
          for a Gaussian $\tem_{00}$ mode.
          The star marks the chosen cavity geometry,
          and subplots depict cuts of the available NA
          \begin{captionenum}
              \item
              at constant $\zr/\lambda$ and
              \item
              at confocality, $L = 2\zr$.
          \end{captionenum}
      \end{captionenum}
  }
  \label{fig:cavity-geometry}
\end{figure}

We report on the design and characterization of a superconducting mm-wave Fabry--P\'erot resonator for use in future atom-array cavity QED experiments.
Motivated by the competing requirements of high cooperativity and optical access,
we adopt a near-confocal geometry (Fig.~\ref{fig:cavity-geometry}).
In this regime,
we find that an understanding of post-paraxial effects
is crucial to avoiding excess loss due to accidental mode hybridization.
Guided by a model of these effects,
we tune the cavity length 
to achieve a finesse~$F=\maxfinesse{}$,
limited primarily by magnetic flux trapping.
The measured finesse corresponds to a cooperativity~$\eta = \etaest$ for long-lived circular Rydberg states.
By establishing the compatibility of a high-cooperativity mm-wave cavity
with optical access for atomic tweezer arrays,
our work paves the way to harnessing strong atom--photon interactions
for nonlocal quantum gates, scalable entanglement, and many-body quantum simulation.

The twin goals of optical access and high cooperativity
place conflicting requirements on the geometry of a Fabry--P\'{e}rot cavity.
This tension is best understood by rewriting the cooperativity in terms of geometric parameters.
For an atom at an antinode at cavity center,
the cooperativity on a cycling transition of wavelength~$\lambda$ is
\begin{equation}
    \label{eq:eta-geom}
    \eta_0 = \frac{6}{\pi^3} \frac{F\lambda^2}{w_0^2}
    =
    \frac{6}{\pi^2} \frac{F \lambda}{\zr},
\end{equation}
where $F$ is the finesse,
$w_0$ the mode waist,
and $\zr = \pi w_0^2/\lambda$ the Rayleigh range~\cite{tanji2011interaction,supp}.
The first expression compares the cross-sectional area of the cavity mode~$\propto w_0^2$ with the atomic cross section~$\propto\lambda^2$,
augmented by the average number~$F/\pi$ of round-trips made by photons in the cavity.
Equivalently, the cooperativity can be written in terms of the solid angle~$\lambda/\zr$ subtended by the cavity mode in the far field which, together with the enhancement factor~$F/\pi$, determines the probability that an emitted photon is scattered into the cavity mode.  Thus, the cooperativity improves as the mode grows more divergent.
A more divergent mode, however, requires cavity mirrors subtending a larger solid angle to maintain a high finesse, limiting the solid angle that remains open for optical access [Fig.~\ref{fig:cavity-geometry}\ref{it:schematic}].


To balance the tradeoff between cooperativity and optical access, we consider the imaging aperture available over the entire design space of cavity length~$L$ and mode size~[Fig.~\ref{fig:cavity-geometry}\ref{it:design-space}], parametrizing the latter by the Rayleigh range~$\zr$. For each combination~$(L, \zr)$, we choose the mirror size such that clipping of the Gaussian mode limits the cavity to a fixed finesse~$F_\text{clip} = 10^{10}$.
We plot the remaining transverse numerical aperture (NA) in Fig.~\ref{fig:cavity-geometry}\ref{it:design-space} as a function of dimensionless parameters~$L/\zr$ and $\zr/\lambda$.
The ratio~$L/\zr$ specifies the cavity geometry and is solely determined by mirror spacing and curvature, with $L/\zr \to 0, \infty$ in the planar and concentric limits,
respectively.  The normalized Rayleigh range~$\zr/\lambda$ is inversely proportional to the mode divergence and thus also to the cooperativity at fixed finesse.

Figure~\ref{fig:cavity-geometry}\ref{it:design-space} shows that the transverse aperture available for imaging improves with increasing cavity length~$L/\zr$ and with increasing Rayleigh range~$\zr/\lambda$, but also shows points of diminishing returns along both axes.
As optical access grows only marginally beyond confocality ($L/\zr = 2$), we opt for a near-confocal geometry,
which additionally provides robustness to misalignment~\cite{siegman1986lasers}.
Here, choosing $z_R/\lambda \approx 6$ provides sufficient optical access ($\mathrm{NA} \approx 0.5$) for tweezer trapping and imaging,
whereas further increasing $z_R/\lambda$
(i.e., decreasing mode divergence)
provides little gain in NA while directly reducing cooperativity.
The resulting cavity design,
depicted in Fig.~\ref{fig:cavity-geometry}\ref{it:cavity-pic},
is marked by the star in Fig.~\ref{fig:cavity-geometry}\ref{it:design-space}.




The cavity consists of two concave toroidal metal mirrors
of diameter~$D = \SI{48}{\mm}$
and curvature radii~$R_{x,y} = \SIlist{42.0;43.5}{\mm}$, with their principal axes aligned.  These mirrors
consist of diamond-machined oxygen-free copper substrates
with a \SI{2}{\micro\meter} film of superconducting niobium~%
\cite{supp}.
We considered three cavity lengths~%
$L_{0,1,2} = \SIlist{43.75;45.44;47.15}{\mm}$;
at \hadrianalen{},
the cavity leaves a numerical aperture~$\mathrm{NA} = \cavityNA{}$ for optical access.
$L_{0,1,2}$ are all slightly longer than confocal
to avoid degeneracy between transverse modes.
We quantify the deviation from confocality
with the parameter~$\mean{g} = 1 - L/\mean{R}$
comparing cavity length
to the harmonic mean mirror curvature radius~$\mean{R}$.
From cavity spectra
in the three geometries,
we determine
$\mean{g}_{0,1,2} = (\numlist{-0.029;-0.068;-0.109})$,
corresponding to $\mean{R} = \SI{42.53}{\mm}$.




We probe the cavity in a closed-cycle dilution refrigerator,
at temperatures~$T$ from \SIrange{0.4}{4}{\K},
using the two coaxial probes shown in Fig.~\ref{fig:cavity-geometry}\ref{it:cavity-pic}.
The probes are oriented at \SI{45}{\degree} from the principal axes of the toroidal mirrors,
permitting coupling to both polarization modes~\cite{supp}.
When one probe is driven at mm-wave frequencies,
cavity modes appear in the resulting transmission~($S_{21}$) spectrum
as narrow resonance features
whose widths give mode finesses
and whose amplitudes quantify the coupling to the probes.
We mount the probes on cryogenic translation stages to adjust this coupling \emph{in situ}
for tuning to a regime where the finesse is limited by intrinsic cavity losses
and not by loss through the probes.


\begin{figure*}[t]
    \centering
    \includegraphics[width=\textwidth]{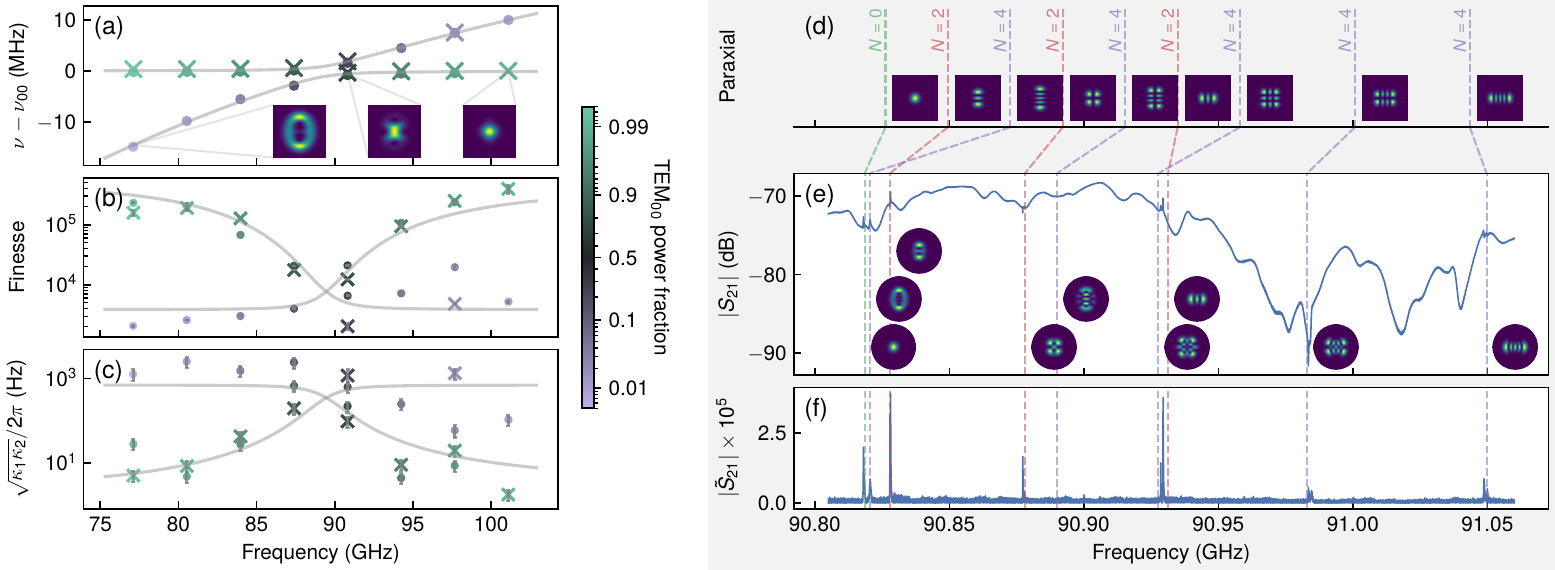}
    \caption[]{%
        Cavity spectroscopy of mode hybridization in the \nearcfcl-geometry, measured at $T = \SI{4}{\K}$.
        \textbf{(a--c)}~%
        Spectroscopy of an avoided crossing between the $\tem_{00}$ mode series and a series of 
        fourth-order modes.
        \begin{captionenum}
            \item\label{it:anticrossing}%
            Mode frequencies~$\nu$
            relative to
            $\tem_{00}$ mode frequencies~$\nu_{00}$
            absent coupling to fourth-order mode,
            \item\label{it:finesse-dip}%
            finesse, and
            \item\label{it:anticrossing-coupling}%
            probe coupling $\sqrt{\kappa_1\kappa_2}$,
            plotted versus absolute mode frequency.
            Points and crosses respectively indicate
            $x$- and $y$-polarized modes.
            Solid gray curves are guides to the eye.
            Insets in \ref{it:anticrossing} show reconstructed transverse intensity profiles for three indicated modes~\cite{supp}.
            \item\label{it:parax}%
            Paraxial prediction for the cavity spectrum around $\tem_{26,0,0}$, with colored dashed lines marking modes of transverse order~$N$, illustrated by transverse intensity profiles.
            \item\label{it:raw-data}%
            Measured cavity $S_{21}$ spectrum
            with mode frequencies predicted by post-paraxial theory~\cite{vanexter2022fine}
            and reconstructed transverse intensity distributions
            in accompanying insets.
            \item
            \label{it:high-passed}%
            Applying a high-pass filter to the raw data in \ref{it:raw-data}
            reveals more clearly the narrow cavity modes,
            with amplitudes dependent on coupling to the probes.
            \end{captionenum}}
    \label{fig:cassia-avoided-crossing}
\end{figure*}

In the geometry closest to confocality, with $\mean{g}_0 = \nearcfclg$, we observed excess cavity losses attributable to mode hybridization. The $\tem_{00}$ mode frequencies [Fig.~\ref{fig:cassia-avoided-crossing}\ref{it:anticrossing}] reveal an avoided crossing with a set of higher-order modes coincident with a dip in $\tem_{00}$ mode finesses  [Fig.~\ref{fig:cassia-avoided-crossing}\ref{it:finesse-dip}, measured at \SI{4}{\K}].
Together, these data suggest that admixture of the transversely larger higher-order mode into the $\tem_{00}$ mode
increases clipping loss at the mirrors.
Measurements of probe coupling [Fig.~\ref{fig:cassia-avoided-crossing}\ref{it:anticrossing-coupling}] support this explanation:
modes with a larger fraction of the higher-order mode couple more strongly to the probes,
which in these measurements were located just outside the edge of the mirrors, \SI{25}{\mm} from the cavity axis.
We quantify probe coupling by
the geometric mean~$\sqrt{\kappa_1\kappa_2}$
of the loss rates of cavity energy through the two probes,
computed from the linewidth and amplitude of the $S_{21}$ response~\cite{supp}.


The intrusion of this higher-order mode
is unexpected in the paraxial theory of Gaussian beams.
Around $\tem_{26,0,0}$, for instance, paraxial theory predicts the spectrum shown in Fig.~\ref{fig:cassia-avoided-crossing}\ref{it:parax},
with $\tem_{26,0,0}$ separated by \SI{25}{\MHz} from the closest higher-order transverse mode---%
far in excess of the \SI{1}{\MHz} coupling to higher-order modes
indicated by the spectral gap in Fig.~\ref{fig:cassia-avoided-crossing}\ref{it:anticrossing}.
However, the cavity operates
in a \emph{quasioptical} regime,
where the wavelength is comparable to characteristic mode dimensions ($\zr/\lambda \lesssim 10$).
The resulting highly divergent modes
are advantageous to cavity QED (Eq.~\ref{eq:eta-geom}),
but necessitate accounting for post-paraxial corrections,
i.e., higher-order terms in the expansion parameter~$1/kw_0 = \sqrt{\lambda/4\pi \zr}$,
where~$k = 2\pi/\lambda$~\cite{lax1975maxwell}.





The consequences of working with such divergent modes
have previously been explored,
primarily in optical microcavities.
Mode mixing there has been found to cause excess losses~ \cite{benedikter2015transverse,podoliak2017harnessing},
though it may also suppress diffractive loss in certain cavity geometries~\cite{kleckner2010diffraction}.
Post-paraxial frequency shifts, too,
have been observed at optical~\cite{uphoff2015frequency,koks2022observation}
and microwave frequencies~\cite{erickson1975high,yu1983high},
and an operator approach has been developed to compute them analytically~\cite{vanexter2022fine}.

Applying the operator approach of Ref.~\cite{vanexter2022fine} to our cavity, we identify the higher-order mode appearing in Fig.~\ref{fig:cassia-avoided-crossing}\ref{it:anticrossing}, showing that post-paraxial effects can explain the surprising mode degeneracies in near-confocal geometries.
The theory predicts mode frequencies
as a function of four geometric parameters---%
cavity length~$L$, mirror curvature radii~$R_{x,y}$, and a coefficient~$\tilde p$ describing the fourth-order curvature of the mirrors---%
which we fit to observed spectra~\cite{supp}.
Figures~\ref{fig:cassia-avoided-crossing}\ref{it:raw-data}\==\ref{it:high-passed}
plot the $S_{21}$ cavity spectrum around $\tem_{26,0,0}$
with model predictions of mode frequencies and mode field patterns,
implicating an errant fourth-order mode in the finesse dip observed in Fig.~\ref{fig:cassia-avoided-crossing}\ref{it:finesse-dip}.
Indeed, the post-paraxial frequency corrections
are comparable to the transverse mode spacings predicted by paraxial theory,
shifting higher-order modes in the \nearcfcl-geometry
much closer to the $\tem_{00}$ modes.

\begin{figure}
    \centering
    \includegraphics[width=\columnwidth]{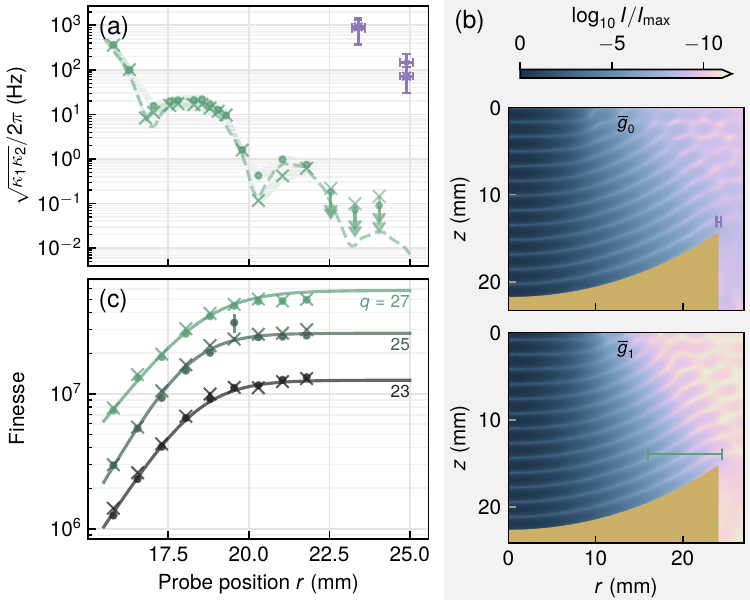}
    \caption[]{%
    Improved localization of cavity modes away from confocality.
    \begin{captionenum}
        \item\label{it:coupling}%
        Probe coupling~$\sqrt{\kappa_1 \kappa_2}$
        versus distance~$r$ of probe tips from cavity axis
        for (purple) $\tem_{26, 0, 0}$ at \SI{90.655}{\GHz}
        in the \nearcfcl{}-geometry
        and for (green)
        $\tem_{27, 0, 0}$ at \SI{90.818}{\GHz}
        in the \lesscfcl{}-geometry.
        Points and crosses show $x$, $y$ mode polarization,
        respectively.
        Bands and error bars
        indicate $1\sigma$ measurement uncertainty,
        while arrows denote upper bounds
        for mode response amplitudes
        below the detection limit.
        Dashed line gives simulation estimates
        in the \lesscfcl-geometry,
        obtained from
        \item\label{it:q3d-sims}%
        finite-element simulations
        of the modes in~\ref{it:coupling},
        where we approximate the cavity as cylindrically symmetric
        for computational efficiency~\cite{supp}.
        We plot electric field intensity
        versus cylindrical coordinates~$(r, z)$,
        normalized to peak intensity.
        Colored lines indicate
        range of positions
        over which probe coupling is plotted in \ref{it:coupling}.
        \item
        \label{it:ringdown-finesse}%
        Finesse of $\tem_{q,0,0}$ modes
        in the \lesscfcl-geometry
        at \SI{0.4}{\K}
        (from dark to light, $q=23, 25, 27$)
        as probes are retracted.
        Curves are guides to the eye.
    \end{captionenum}
    }
    \label{fig:coupling-strength}
\end{figure}

To avoid mode mixing, we lengthened the cavity to a geometry further from confocal, with $\lesscfcl{} = \lesscfclgnum{}$.
Here, the higher-order transverse modes are pushed to spacings~$\gtrsim\SI{100}{\MHz}$,
much greater than the $\SI{1}{\MHz}$ coupling strength
observed in the \nearcfcl{}-geometry,
so the $\tem_{00}$ modes
should be purer 
and consequently better localized along the cavity axis.
Figure~\ref{fig:coupling-strength}\ref{it:coupling}
compares probe coupling~$\sqrt{\kappa_1 \kappa_2}$
as a function of probe position
for two analogous $\tem_{00}$ modes in the \nearcfcl{}- and \lesscfcl{}-geometries.
As the probes are retracted
in the \lesscfcl{}-geometry,
this coupling drops steeply
to levels far below those observed in the \nearcfcl{}-geometry.
The measured couplings, directly proportional to the local mode field intensity at the probes,
agree with the simulated intensity profiles
plotted in Fig.~\ref{fig:coupling-strength}\ref{it:q3d-sims}~%
\cite{supp,marsic2018modal,schnaubelt2021comparison},
confirming the improved mode localization in the \lesscfcl{}-geometry.



We confirm that the reduction in mode mixing away from confocality translates into improved finesse
by performing cavity ringdown spectroscopy in the \lesscfcl{}-geometry
and in an even longer cavity with $\hadriana{} = \hadrianagnum{}$.
The ringdown measurement avoids broadening of the cavity line due to sub-\si{\nm} oscillations of the cavity length driven by residual vibrations in the cryostat~\cite{supp}. We further operate at a lower temperature~$T=\SI{0.4}{\K}$ to avoid being limited by temperature-dependent cavity loss.  Figure~\ref{fig:coupling-strength}\ref{it:ringdown-finesse} shows the finesse of three $\tem_{00}$ modes in the \lesscfcl{}-geometry as a function of probe position.  The finesse initially increases as the probes are retracted, consistent with their decreased coupling to the cavity, before plateauing to unloaded finesses~$F\gtrsim 10^7$, a factor $\sim 10^2$ higher than achieved in the \nearcfcl-geometry.

\begin{figure}[t]
    \centering
    \includegraphics[width=\columnwidth]{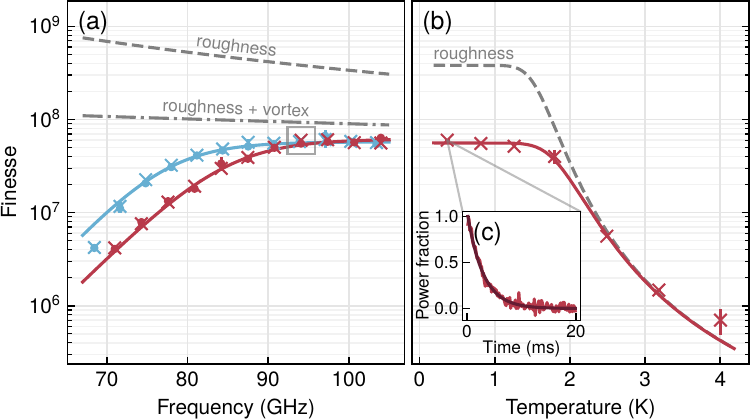}
    \caption[]{%
    Cavity performance.
    \begin{captionenum}
        \item\label{it:best-fins}%
        Finesse of $\tem_{00}$ cavity modes
        at \SI{0.4}{\K} in the \lesscfcl{}- (\geminacolor{})
        and \hadriana{}- (\hadrianacolor{}) geometries
        compared
        to limits set by mirror roughness and magnetic flux trapping.
        (Points, crosses: $x$- and $y$-polarized modes,
        respectively.)
        Solid lines are guides to the eye.
        For the $\tem_{28,0,0,y}$ mode
        of the \lesscfcl{}-geometry
        at \SI{94.073}{\GHz}
        (\geminacolor{} cross, boxed),
        we show
        \item\label{it:temp-dep}%
        finesse versus cavity temperature.
        Solid line fits the temperature dependence
        as a combination of
        superconductor ac resistance
        and a constant residual loss~%
        \cite{supp};
        dashed line shows the same fit
        with the lower residual loss
        set by the roughness limit in \ref{it:best-fins}.
        \item
        \label{it:ringdown-inset}
        Ringdown data behind the measurement at \SI{0.4}{\K};
        $\kappa = 2\pi\times\SI{55.1(1.5)}{\Hz}$ decay of the fitted exponential curve
        yields finesse~$F = \num{5.99(16)e+7}$.%
    \end{captionenum}%
    }
    \label{fig:cavity-perf}
\end{figure}


Figure~\ref{fig:cavity-perf}\ref{it:best-fins} summarizes the improved finesse as a function of mode frequency in the \lesscfcl{}-geometry (\geminacolor{} markers) and the longer \hadriana{}-geometry (\hadrianacolor{} markers).  For both cavity lengths, the finesse of the $\tem_{00}$ modes increases with increasing frequency~$\nu$ before plateauing to a common value~$\Fmax = \maxfinesse{}$ for $\nu \gtrsim \SI{90}{\GHz}$. We attribute the lower finesses at low frequencies to residual mode mixing, which is more significant for the more confocal (\lesscfcl{}) geometry. The plateau to the same finesse~$\Fmax$ for both geometries at high frequency indicates that mode mixing is no longer a dominant limitation in this regime. Rather, the remaining losses at high frequency are well explained by magnetic flux trapping and a smaller contribution from the surface roughness of the mirrors.

Both residual mode mixing and magnetic flux trapping
can be mitigated in future experiments to reach even higher finesse.
Further detuning the cavity away from confocality should improve the purity of the $\tem_{00}$ modes,
and hence their transverse localization,
to reduce the diffractive losses we observe at low frequencies.
Additionally, mirror profiles may be numerically optimized
to suppress mode mixing
using the finite-element simulations  
shown in Fig.~\ref{fig:coupling-strength}\ref{it:q3d-sims},
which provide fast, reliable estimates of even very low mode losses~%
\cite{marsic2018modal,schnaubelt2021comparison,supp}.
At high frequencies, the dominant source of loss is flux trapping due to an ambient magnetic field with component~$B_\perp = \SI{0.22}{\gauss}$ normal to the mirrors, as shown in Fig.~\ref{fig:cavity-perf}\ref{it:best-fins} (dot-dashed curve). Suppressing the magnetic field during cavity cooldown to $B_\perp \lesssim \SI{10}{\milli\gauss}$
would yield high-frequency cavity performance limited only by the mirror surface roughness (dashed curve).
The measured roughness~$h_\text{rms} = \SI{23}{\nm}$ sets a finesse limit~$F_\text{surf} = \pi/4k^2 h_\text{rms}^2 \gtrsim \num{3e+8}$.

Crucial for achieving high finesse is operating below the superconducting transition temperature~$T_c = \SI{9.2}{\K}$ of the niobium mirror coatings.
Even in this regime, BCS theory predicts a residual temperature-dependent ac resistance.  To understand the implications for the required operating temperature, we measured the finesse in the \lesscfcl-geometry for temperatures~$T$ from \SIrange{0.4}{4}{\K}, as shown in Fig.~\ref{fig:cavity-perf}\ref{it:temp-dep} for the $\tem_{28,0,0,y}$ mode.
We observe a finesse consistent with BCS resistance, roughly following a Boltzmann scaling~$R_\text{BCS}(T) \sim e^{-\Delta/k_B T}$ governed by the superconducting gap~$\Delta/k_B = \SI{17.67}{\K}$,
until reaching the limiting value~$\Fmax$ at low temperature.
Encouragingly, the cavity can operate at temperatures up to \SI{1.3}{\K}
without impacting cavity performance,
even if the finesse were limited only by surface roughness.
This flexibility is important for cryogenic atom array experiments,
where heat loads due to scattered trapping light may limit the achievable temperature.

The cavity is poised to benefit from ongoing advances in developing cryogenic atom arrays~\cite{schymik2021single,zhang2025high,pichard2024rearrangement} and controlling circular Rydberg states.
Circular Rydberg atoms, recently prepared in tweezer arrays~\cite{mehaignerie2025interacting,wirth2024quadrupole},
offer long radiative lifetimes at low temperature
and maximal cooperativity,
as their only dipole-allowed radiative transition
is the mm-wave transition that couples to the cavity.
Coupling the \SI{92}{\GHz} transition
between circular Rydberg states $\ket{41C}\leftrightarrow\ket{42C}$ to a linearly polarized cavity mode
yields a cooperativity~$\eta = \eta_0 / 2 = \etaest$.  The corresponding parameters~$(g, \kappa, \Gamma) = 2\pi \times \SIlist{22e+3;55;13}{\Hz}$
permit nonlocal entangling gates~\cite{zheng2000efficient,sarkany2015long,sorensen2003measurement,jandura2024nonlocal}
with fidelity~$\gtrsim \SI{98}{\percent}$~\cite{supp}.


The prospect for deterministic mm-wave--mediated entanglement in atom arrays
opens myriad opportunities in quantum information processing,
from manipulating nonlocally encoded information in topologically ordered states~\cite{jiang2008anyonic}
to investigating low-overhead quantum error correcting codes
that rely on nonlocal connectivity~\cite{bravyi2010tradeoffs, baspin2022quantifying,bravyi2024high}.
The cavity also offers a route
to scalably generating non-classical states
of atoms or photons
as metrological resources,
e.g., for Heisenberg-limited phase estimation~\cite{monz201114,davis2016approaching}.
Further, the interplay of cavity-mediated nonlocal interactions with local dipolar interactions
opens rich possibilities in quantum many-body physics,
from inducing maximally chaotic dynamics~\cite{belyansky2020minimal}
to stabilizing quantum spin liquids~\cite{chiocchetta2021cavity}.

\paragraph*{Acknowledgments.}
We thank May Ling Ng for her assistance
with mirror surface profilometry,
and acknowledge
S.~Kuhr, S.~Gleyzes,
C.~Koks, M.~van Exter,
N.~Marsic, E.~Schnaubelt, and J.~Simon
for helpful discussions.  This work was supported by the U.S. Department of Energy (DOE) Office of Science, Office of High Energy Physics and Q-NEXT National Quantum Information Science Research Center under contract number DE-AC02-76SF00515.  We additionally acknowledge support from the Office of Naval Research under award No.~N00014-21-1-2451 (M.\,W., M.\,S.-S.), the Gordon and Betty Moore Foundation (M.\,W.), and the DOE Laboratory Directed Research and Development program at SLAC National Accelerator Laboratory under contract DE-AC02-76SF00515 (L.\,X.).  T.\,Z.\ was supported by the DOE
Office of Science Graduate Student Research (SCGSR) program,
administered by the Oak Ridge Institute for Science and Education
under contract number DE‐SC0014664.  S.\,R.\,C.\ acknowledges support from the National Science Foundation Graduate Research Fellowship program.  K.\,K.\,S.\,M.\ acknowledges support from the Natural Sciences and Engineering Research Council of Canada (NSERC).
N.\,P.\ acknowledges support from the Hertz Foundation.
A.-M.\,V.-F. is supported by the DOE Office of Science, Office of Nuclear Physics,
under contract DE-AC05-06OR23177.



\bibliography{aipsamp}

\clearpage

\supplemental{}

This supplemental material provides additional technical details and theory in support of the main text. Section~\ref{sec:physical} describes cavity mirror fabrication and the measurement apparatus used to probe the cavity. We turn to supporting theory for our interpretation of cavity spectra (Sec.~\ref{sec:spectrumtheory}) and for our measurements of the finesse (Sec.~\ref{sec:measurementtheory}). Section~\ref{sec:sims} provides technical details behind the finite-element simulations of cavity modes shown in Fig.~\ref{fig:coupling-strength},
while
Sec.~\ref{sec:losses} presents the theory
behind our modeling of cavity loss.
Finally, Sec.~\ref{sec:atoms-numbers} relates cavity performance
to the coherence of atom--light interactions,
computing the projected cooperativity
and offering a concrete protocol for an entangling gate
attaining the fidelity quoted in the main text.

\section{Apparatus}
\label{sec:physical}

\subsection{Mirror fabrication}
\label{sec:mirror-fab}

As described in the main text,
our cavity consists of a pair of copper mirrors
coated with a thin film of superconducting niobium.
This niobium-on-copper approach reduces diffractive cavity loss,
but introduces some complexity in fabrication.
Thin film niobium provides superior surface roughness compared to
superconducting rf cavities in accelerator applications,
which are directly machined from bulk niobium.
This advantage stems from the relative ease of achieving low surface 
roughness in a substrate material.
In contrast to accelerator cavities,
low surface roughness is critical for our cavity
due to the potential for diffractive losses with an open cavity geometry
and our higher operating frequencies.

\begin{figure}[t]
    \centering
    \includegraphics[width=0.65\textwidth]{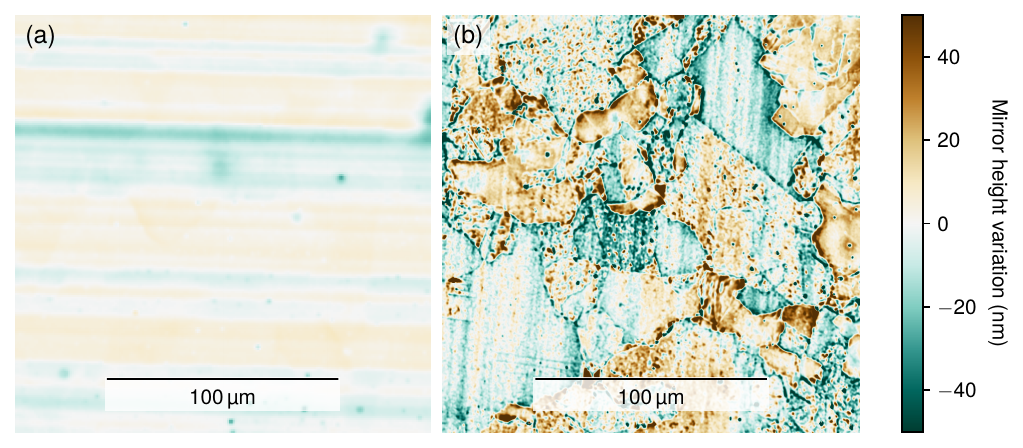}
    \caption[]{%
    Interferometric measurements
    of cavity mirror surface roughness.
    Large-scale height variations
    are removed by a fit to a paraboloid
    such that fit residuals reveal small-scale surface roughness.
    \begin{captionenum}
        \item\label{it:substrate}%
        The diamond-machined copper substrates
        attain an rms roughness~$h_\text{rms} = \SI{5}{\nm}$.
        \item\label{it:coated}%
        After niobium deposition,
        the surface roughness increases to $h_\text{rms} = \SI{23}{\nm}$.
    \end{captionenum}}
    \label{fig:mirror-roughness}
\end{figure}

Fabrication begins with the preparation of copper mirror substrates.
Oxygen-free copper blanks are roughly machined
and then vacuum-baked at \SI{350}{\celsius} for \SI{5}{\minute}
for stress relief.
These mirror blanks subsequently undergo diamond machining
to achieve an rms surface roughness~$h_\text{rms} = \SI{5}{\nm}$
(KAF Manufacturing),
as shown in Fig.~\ref{fig:mirror-roughness}\ref{it:substrate}.

The substrates are coated with a \SI{2}{\um} layer of niobium
at the Thomas Jefferson National Accelerator Facility.
We choose a thickness
far exceeding the penetration depth~$\london \approx \SI{40}{\nm}$
of niobium
so that incident radiation does not interact with the copper substrate.
To prepare the substrates for deposition,
they are degreased,
passivated with sulfamic acid,
and rinsed with methanol.
After a 24 hour bake-out at \SI{150}{\celsius},
we deposit the niobium by energetic condensation
with an electron cyclotron resonance plasma~\cite{valente2014development,valente2016superconducting},
with \SI{184}{\eV} ion energy to nucleate the film
followed by subsequent growth at \SI{64}{\eV}.
Halfway through the subsequent growth at \SI{64}{\eV},
the deposition is interrupted for a few hours.
The interrupted deposition method has demonstrated
lower surface resistance
both for niobium films deposited via energetic condensation
and for those obtained from DC magnetron sputtering.
The substrates are maintained at \SI{150}{\celsius}
throughout the process.
This temperature is sufficiently low
to prevent surface roughening due to Ostwald ripening
while still ensuring sufficient mobility for the niobium adatoms.

The resulting niobium films are markedly rougher than the substrate,
with $h_\text{rms} = \SI{23}{\nm}$.
This roughness is dominated by
the differential growth of niobium grains,
visible in Fig.~\ref{fig:mirror-roughness}\ref{it:coated}.
We shall be interested in the normal-state resistivity~$\rho_n$
of the films at cryogenic temperatures
when we turn to estimate cavity losses
(Sec.~\ref{sec:losses}).
Therefore, we perform four-point probe measurements
of the dc resistivity
of witness samples deposited on $\mb a$-plane sapphire
just above the superconducting transition temperature~$T_c$.
The two samples corresponding to the two cavity mirrors
have
residual resistivity ratios~$\RRR = \rho_{\SI{300}{\K}}/\rho_n = \numlist{52;57}$,
where $\rho_{\SI{300}{\K}} = \SI{152}{\nano\ohm\meter}$~%
is the room-temperature resistivity of niobium.
RRR typically ranges from \numrange{30}{300}
in niobium used for superconducting~rf applications.

\subsection{Cryogenics and rf}
\label{sec:apparatus}

\begin{figure}[h]
    \centering
    \includegraphics[width=\textwidth]{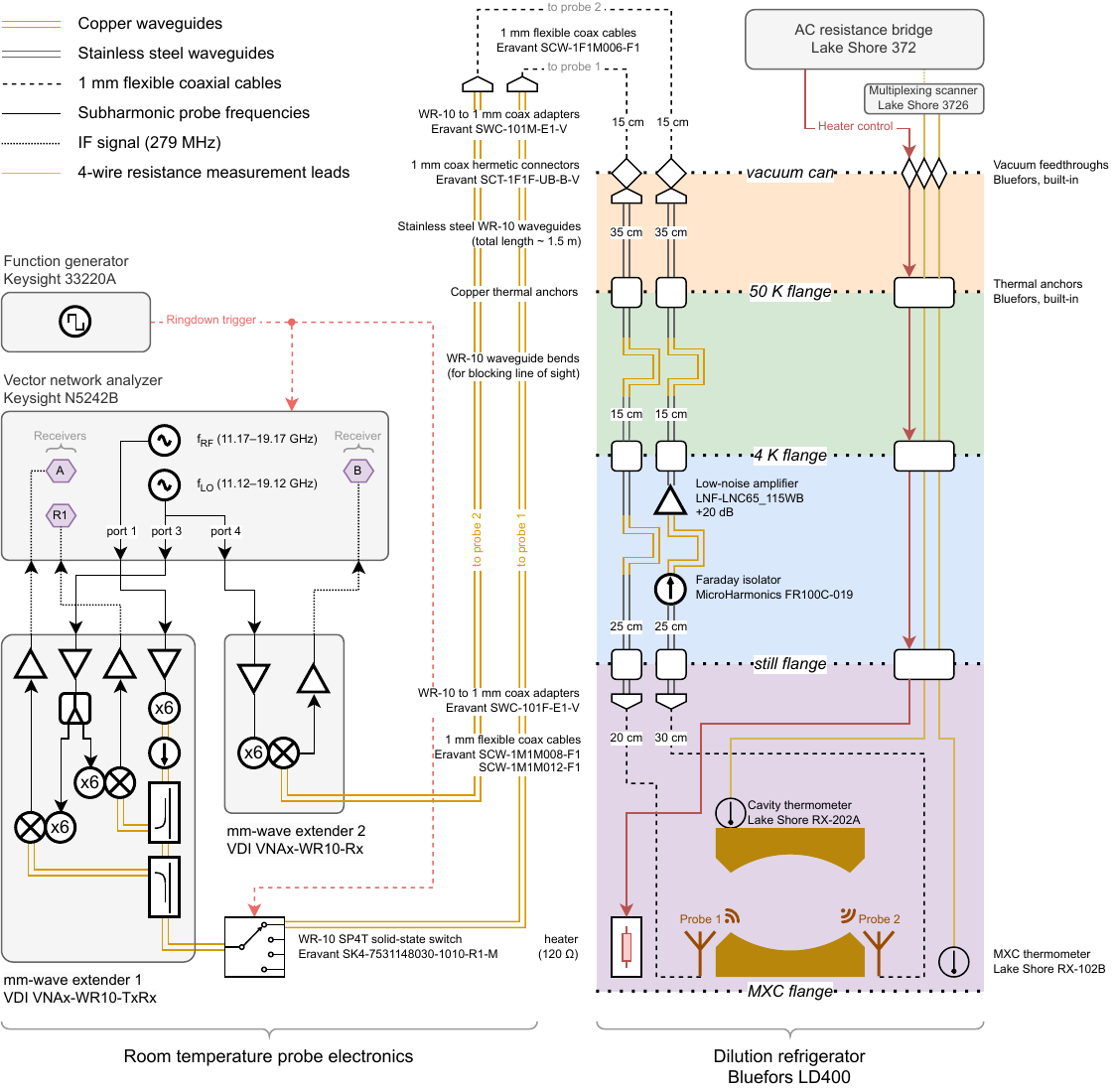}
    \caption[]{Block diagram of key experimental components.
    Broadly, the apparatus consists of room temperature probe electronics
    connected to the cavity in the dilution refrigerator
    by long lengths of stainless steel waveguides
    for thermal isolation.
    At room temperature,
    a vector network analyzer (VNA)
    is connected to two mm-wave extenders 
    that up-convert a microwave drive tone to mm-wave frequencies
    and perform heterodyne detection of incoming mm-waves,
    downmixing them to \SI{279}{\MHz} signals for the VNA.
    A switch at the output of extender 1
    allows for ringdown measurements,
    during which a function generator controls the switch
    and triggers the VNA.
    Inside the cryostat,
    a chain of waveguides, coaxial cables, and adapters
    connect the extenders to the cavity probes.
    A cryogenic low-noise amplifier boosts signal levels for $S_{21}$~measurements,
    partly compensating for the large insertion loss
    of the stainless steel waveguides.
    Two thermometers and a heater provide temperature monitoring and control.
    }
    \label{fig:block-diagram}
\end{figure}

Characterizing the superconducting cavity
requires a cryogenic apparatus capable
of stable operation at temperatures
much lower than the transition temperature~$T_c = \SI{9.2}{\K}$
of the niobium mirror coatings,
together with electronics for probing the cavity
at millimeter-wave frequencies.
We describe here the experimental setup in detail
and present a block diagram in Fig.~\ref{fig:block-diagram}.

As outlined in the main text,
all measurements are conducted in a closed-cycle dilution refrigerator
(Bluefors LD400)
with the cavity typically mounted at the mixing chamber (MXC) flange.
At the lowest temperatures used in this work,
from \SIrange{0.3}{0.8}{\K},
we operate the cryostat with dilution cooling,
while for measurements from \SIrange{0.8}{4}{\K},
we operate the cryostat without \ce{^3He}
and rely only on evaporative cooling of \ce{^4He}.
In either case,
we maintain stable experimental temperatures
by feedback control of a resistive heater,
monitoring temperature with
a built-in ruthenium oxide sensor mounted on the flange.
Closed-loop control is essential in evaporative mode:
we cannot regulate the helium vapor pressure in the cryostat,
so the circulation pump pumping speed
sets an upper bound $\approx \SI{1.3}{\K}$
above which no steady-state exists without active feedback.

To ensure that the cavity is well thermalized to the cryostat,
we affix the cavity mirrors to a mount made of oxygen-free copper
and clamp the full cavity block onto the MXC flange,
which itself consists of a copper plate
coated with gold to reduce thermal contact resistance.
Even so, we are concerned with the possibility of thermal gradients
across the apparatus
distorting the data of Fig.~\ref{fig:cavity-perf}\ref{it:temp-dep},
given the dissipation of mm-wave power during our measurements.
Thus, we additionally monitor the temperature at the top of the cavity block
with a second ruthenium oxide sensor
calibrated off the built-in flange sensor.
Only below \SI{1}{\K}
do decreasing cryostat cooling power
and thermal conductivity in the mount
start producing discernible gradients
between the top of the cavity
and the MXC flange
(\SI{20}{\milli\K} at \SI{0.8}{\K}
and \SI{100}{\milli\K} at \SI{0.3}{\K}).

We characterize the cavity with two coaxial probes.
Each probe consists of
a length of \SI{1.2}{\mm}--diameter
semi-rigid coaxial cable
with \SI{2}{\mm} of
the \SI{0.29}{\mm}--diameter center conductor
exposed at the tip.
The probes couple to the electric field of cavity modes
through this exposed tip~%
(Sec.~\ref{sec:probe-coupling}).
These probes are positioned
to provide robust, tunable, and symmetric coupling
to all cavity modes of interest.
We orient the probes radially inward
on a plane \SI{45}{\degree}
from the principal axes of the toroidal mirrors,
permitting coupling to modes polarized
along either principal axis.
To maximize coupling to cavity modes,
we keep the probes near and roughly parallel
to the surface of one mirror.
since the modes are largest transversely
at either end of the cavity
with electric fields parallel to the mirrors.
(The probe tips are then approximately orthogonal to each other,
minimizing direct crosstalk between probes
by cross-polarizing their dipole radiation.)
We tune probe coupling
using cryogenic translation stages (JPE CBS10-RLS),
maintaining equal coupling for the two probes
by moving the stages in tandem
to maintain symmetric probe positioning.
The stages give a tuning range of \SIrange{15.5}{25.0}{\mm}
for the distance~$r$ between the probe tips
and the cavity axis.

At room temperature,
a vector network analyzer
and two mm-wave extension modules
synthesize and detect our mm-wave signals
between \SIlist[list-pair-separator={ and }, list-units=single]{67;115}{\GHz}.
The extension modules consist of a transceiver (``module 1'')
and a receiver unit (``module 2'')
that up- and down-convert between microwave and mm-wave frequencies
to allow $S_{11}$ and $S_{21}$ measurements.
A switch at the mm-wave output port of module 1
allows us to shut off probe power
for ringdown measurements.
We connect the test ports to the two cavity probes
(1 and 2, corresponding to our module labeling)
via WR10 stainless steel rectangular waveguides
with a total length of \SI{1.5}{\meter}.
These stainless steel waveguides reduce heat load on the cryostat
at the cost of high insertion loss~(\SI{-40}{\dB} total).
We compensate for this loss
with a low-noise amplifier at \SI{4}{\K}
(\SI[retain-explicit-plus]{+20}{\dB})
on the signal exiting from probe 2.

We pay particular attention
to the vibration environment of the cavity,
whose narrow mode linewidths
make measurements extremely sensitive to
small variations in the cavity frequency.
A rigid cavity mount
limits differential motion of the two cavity mirrors,
and active vibration stabilization in the frame of the cryostat
isolates the apparatus
from external laboratory vibrations.
The cryostat itself generates additional vibrations in normal operation:
a pulse tube cryocooler providing cooling down to \SI{4}{\K}
contributes a broadband vibration background,
while helium circulation turbopumps
produce significant vibrations
at their rotational frequency of \SI{820}{\Hz}
and at harmonics thereof.
We therefore disable the pulse tube
temporarily during measurements,
and operate entirely without the helium circulation turbopump
at the expense of cooling power and attainable experimental temperature.
Even with these precautions,
we still observe residual vibrations
[Figs.~\ref{fig:q-circles}\ref{it:q-circle-data} and \ref{fig:ringdowns}],
which we handle by taking vibration-insensitive ringdown measurements~%
(Sec.~\ref{sec:ringdown}).

\afterpage{\clearpage}

\section{Cavity spectrum theory}
\label{sec:spectrumtheory}

\subsection{Post-paraxial frequency corrections}
\label{sec:postparax}

This work makes extensive use of the post-paraxial theory
presented by van Exter \emph{et al.}\ in Ref.~\cite{vanexter2022fine},
which we briefly sketch here.
We use theoretical predictions of mode frequencies
to characterize the cavity geometries spectroscopically
and to positively identify our $\tem_{00}$ modes (Sec.~\ref{sec:mode-id}).
As mentioned in the main text,
these predictions also identify the mode
involved in the avoided crossing observed
in the \nearcfcl{}-geometry closest to confocality
(Fig.~\ref{fig:cassia-avoided-crossing})
and motivate our detuning further from confocality
in the \lesscfcl{}- and \hadriana{}-geometries.



\newcommand\hfull{\ensuremath{\hat{H}_\text{phase}}}
\newcommand\etaastig{\ensuremath{\eta_\text{astig}}}

We consider two-dimensional electric field cross-sections
in planes orthogonal to the cavity axis
and their evolution as they propagate around the cavity.
We will view the Hilbert space of such cross-sections
as a tensor product
between the space of wavefunctions in two dimensions
(encoding the mode intensity profiles)
and $\mathbb{C}^2$
(parametrizing the polarization degree of freedom
using Jones vectors).
We then define an operator~$\hat M$
giving the effect of a full trip around the cavity
starting from one of the cavity mirrors.
Assuming no losses,
this round-trip operator~$\hat M$ is unitary
and may be expressed as an exponential~$e^{i\hfull}$
with a Hermitian operator~$\hfull$
giving the phase advance of a mode profile
over a full cycle around the cavity.

To find the eigenmodes of the cavity,
we look for eigenvectors of~$\hfull$
with eigenvalues~$2\pi q$ for integer~$q$,
corresponding to field profiles that accrue $2\pi q$ phase.
These eigenvectors are then fixed points of the round-trip operator~$\hat M$,
as expected for a resonant cavity mode.
Let us first remove from $\hfull$ the trivial phase accrual
from plane wave propagation:
\begin{equation}
    \label{eq:hfull}
    \hfull = 2kL - \hat H.
\end{equation}
In this expression,
$k$ is the wavenumber
(known approximately from paraxial theory),
$L$ the cavity length,
and the operator~$\hat H$ then captures
deviations from plane wave propagation.
The eigenvectors and eigenvalues of this operator~$\hat H$
directly yield the resonant modes of the cavity
and their frequencies:
if $\hat H\ket\psi = \phi\ket\psi$,
then setting $k = \frac{2\pi q + \phi}{2L}$ shows that $\ket\psi$
is an eigenmode of the cavity with frequency
\begin{equation}
    \label{eq:post-parax-freq}
    \omega = ck = \underbrace{\frac{2\pi c}{2L}}_{\omega_\text{FSR}} \left(q + \frac{\phi}{2\pi}\right),
\end{equation}
where $c$ is the speed of light in vacuum
and where we have identified the cavity free spectral range~$\omega_\text{FSR}$.
We shall see that $\hat H$ itself contains a $k$-dependence~$\hat H(k)$,
so in practice we set $k$
to the paraxial frequency
of a mode of interest,
then diagonalize the corresponding $\hat H(k)$ (including post-paraxial effects)
to yield an eigenvalue~$\phi$
differing only slightly from the paraxial value.
The mode frequency calculated
from Eq.~\ref{eq:post-parax-freq}
is thus a self-consistent estimate.
It remains to find the form of $\hat H(k)$
and diagonalize it to find the frequencies and eigenmodes of the cavity.

We shall see that the operator~$\hat H(k)$ consists of a sum of terms
each describing a different effect.
These largely arise from either
the position-dependent phase shift induced by the cavity mirrors
or the deviation from plane-wave propagation
for localized field distributions.
A non-planar cavity mirror can be viewed as a position dependent phase shift~$2k\Delta z$,
where $\Delta z$ is the deviation of the mirror from some reference surface,
conventionally a plane at the mirror apex,
and where the extra factor of two is incurred
because the deviation~$\Delta z$ affects the optical path length
of both the incident and reflected light.
Meanwhile, deviations from plane wave propagation
affect the phase accrued as the mode propagates.
The round-trip operator~$\hat M$
is composed in part by operators~$e^{i \hat k_z L}$
giving the effect of optical propagation over distance~$L$,
where $\hat k_z$~is the axial component of the wavevector.
Writing~$\hat k_z$ in terms of the transverse photon momentum~$\hat{\mb k}_\perp$ as
\begin{equation}
    \label{eq:kz}
    \hat k_z = \sqrt{k^2 - \hat{\mb k}_\perp^2} = k - \frac{\hat{\mb k}_\perp^2}{2k} - \frac{\hat{\mb k}_\perp^4}{8k^3} + O(\hat{\mb k}_\perp^6),
\end{equation}
we identify a constant term~$k$ giving the plane-wave phase accrual,
which we removed in Eq.~\ref{eq:hfull},
as well as a quadratic and quartic term,
which we shall see respectively describe paraxial and (leading-order) post-paraxial propagation.

Let us now consider the actual form of~$\hat H(k)$.
Under the paraxial approximation,
$\hat H(k)$~reduces to a 2D quantum harmonic oscillator Hamiltonian~$\hat H_\text{parax}$,
illustrating the well-known isomorphism
between paraxial optical propagation,
with its Hermite-Gaussian (HG) or Laguerre-Gaussian (LG) modes equispaced in frequency,
and the quantum harmonic oscillator,
which has a ladder of similarly shaped eigenmodes
with equal energy spacings.
Intuitively, this harmonic oscillator Hamiltonian
arises from the leading-order quadratic shape of the cavity mirrors,
which produces a Hamiltonian term of the form $\hat{\mb r}^2$,
and from the paraxial $\hat{\mb k}_\perp^2$~term
in the expansion of~$\hat k_z$ (Eq.~\ref{eq:kz}).
From the paraxial theory of symmetric Fabry--P\'erot cavities~\cite{siegman1986lasers},
we have
\begin{equation}
    \hat H_\text{parax}
    = 2\cos\inv\mean{g} \, (\hat a_x^\dagger \hat a_x + \hat a_y^\dagger \hat a_y + 1)
    = 2\cos\inv\mean{g} \, (\hat a_+^\dagger \hat a_+ + \hat a_-^\dagger \hat a_- + 1),
\end{equation}
where parameter~$\mean{g} = 1 - L/\mean{R}$
compares cavity length~$L$
and mirror curvature radius~$\mean{R}$,
and where we have introduced ladder operators
\begin{align}
    \hat a_x
    &= \frac{\gamma\inv \hat x + i\gamma \hat{k}_x}{\sqrt{2}},
    &
    \hat a_y &= \frac{\gamma\inv \hat y + i\gamma \hat{k}_y}{\sqrt{2}},
    &
    a_\pm &= \frac{\hat a_x \mp i \hat a_y}{\sqrt2}
\end{align}
using the position operators~$\hat x, \hat y$
and momentum operators~$\hat k_{x,y} = -i\hat\pa_{x,y}$.
The Cartesian lowering operators~$\hat a_{x,y}$
(and circular lowering operators~$\hat a_\pm$)
remove excitations from HG (resp.\ LG) mode profiles,
and are defined in terms of
a characteristic length scale~$\gamma = w_1/\sqrt{2}$
set by $w_1$, the $e^{-2}$~spot size of Gaussian modes at mirror apex.

These ladder operators are exactly those of the 2D quantum harmonic oscillator,
so we also define number operators
giving the number of transverse excitations in a mode.
For HG modes,
the eigenvalues of $\hat n_{x,y} = \hat a_{x,y}^\dagger \hat a_{x,y}$
give the two transverse indices,
while for LG modes,
the number operators~$\hat n_\pm = \hat a_\pm^\dagger \hat a_\pm$
give eigenvalues~$n_\pm$ related to LG indices~$p,\ell$
as $p = \min(n_\pm)$ and $\ell = n_+ - n_-$.
Adding the number operators
gives the total transverse mode order~%
$\hat N = \hat n_x + \hat n_y = \hat n_+ + \hat n_-$.

Expressing corrections due to leading-order post-paraxial effects
and to astigmatism
using the harmonic oscillator algebra
then yields the following Hamiltonian~$\hat H$
for a symmetric Fabry--P\'erot cavity:
\begin{multline}
    \label{eq:hterms}
    \hat H = \hat H_\text{parax}
    + \underbrace{\frac{\alpha^2 \gamma^4 \hat{\mb k}_\perp^4}{2 k \mean{R}}}_{\hat H_\text{prop}}
    + \underbrace{\frac{(3 - \alpha^2)\gamma^{-4}\hat{\mb r}^4 - 4(\hat N+1)^2}{2 k \mean{R}}}_{\hat H_\text{wave}}
    + \underbrace{\frac{(1 - \alpha^2)\tilde p \gamma^{-4}\hat{\mb r}^4}{2 k \mean{R}}}_{\hat H_\text{asphere}}
    +
    \underbrace{\frac{-2}{k\mean{R}} (1 + \hat L_z \otimes \hat S_z)}_{\hat H_\text{vec}}
    \\
    + \underbrace{2\etaastig \sqrt{\alpha^2 - 1} \gamma^{-2}(\hat x^2 - \hat y^2)}_{\hat H_\text{astig}}
    +
    \underbrace{\frac{-2\etaastig \hat S_x}{k \mean{R}}}_{\hat H_\text{v+a}}.
\end{multline}
In this expression,
$\alpha = w_1/w_0$ is the expansion in a Gaussian beam
from the waist to a mirror,
aspheric coefficient~$\tilde p$
linearly parametrizes the fourth-order curvature of the mirrors
(with $\tilde p = 0$ for a sphere
and $\tilde p = 1$ for a paraboloid),
$\etaastig = \frac{R_y - R_x}{R_y + R_x}$
quantifies the astigmatism of the mirrors,
$\hat L_z = \hat n_+ - \hat n_-$ is the orbital angular momentum,
and $S_x=\ket{R}\bra{L} + \mathrm{h.a.}$
and $S_z=\ket{R}\bra{R} - \ket{L}\bra{L}$
are Stokes operators
on the space of Jones vectors
(where $R$ and $L$ respectively denote right- and left-handed circular polarization).
Equation~\ref{eq:hterms}
neglects coupling between modes of differing transverse order~$N$
by dropping excitation-nonconserving products of ladder operators
(e.g.\ $a_+^\dagger a_-^\dagger$, $a_+^2$).
Such an assumption is justified
as long as the cavity
is not near-planar ($\mean{g} \not\approx 1$),
ensuring that modes of equal longitudinal order~$q$
but differing transverse order~$N$
are far detuned from each other.

The individual terms correspond to distinct corrections
to the paraxial description of Fabry--P\'erot cavities.
Propagation correction~$\hat H_\text{prop}$
gives the effect of the fourth-order term in Eq.~\ref{eq:kz},
while the wavefront corrections~%
$\hat H_\text{wave}$ and $\hat H_\text{asphere}$
together account for the phase shift
due to mismatch between the cavity mirror profile
and the (fourth-order) shape of Gaussian mode wavefronts.
The vector shift~$\hat H_\text{vec}$ is a spin--orbit coupling
arising from small axial electric fields
not considered in the (scalar) theory of paraxial optics.
Finally, in astigmatic cavities,
the astigmatic part of the mirror profile
produces an extra wavefront correction~$\hat H_\text{astig}$
and an anisotropic spin--orbit coupling~$H_\text{v+a}$.

Given operator~$\hat H$,
we may find the resonant frequencies
and eigenmodes of the cavity
by direct diagonalization.
Alternatively, we may view $\hat H$
as a model allowing us to spectroscopically characterize cavity geometry
by fitting predictions of mode frequencies
to their observed values,
as described in the next section.
Despite the apparently complex form of~$\hat H$ in Eq.~\ref{eq:hterms},
there are only four free parameters:
cavity length~$L$,
harmonic mean mirror curvature radius~$\mean{R}$,
astigmatism parameter~$\etaastig$,
and aspheric parameter~$\tilde p$.
The $\tem_{00}$ modes, for example,
have frequencies
\begin{equation}
    \label{eq:post-parax-00}
    \omega_{q, 0, 0, \{x, y\}} =
    \omega_\text{FSR}\left(
      q
      + \frac{\cos\inv \mean{g}}{\pi}
      - \frac{1 + \tilde p \frac{L}{2\mean{R} - L} \pm 2\etaastig}{2\pi k \mean{R}} \right),
\end{equation}
where the first two terms are well-known from the paraxial theory
and the final term is the post-paraxial correction,
which includes a splitting between $x$- and $y$-polarized modes
due to astigmatism.

\subsection{Mode identification}
\label{sec:mode-id}

In this section,
we outline our approach
for identifying the $\tem_{00}$ modes of interest
in the $S_{21}$ spectra of the cavity.
Post-paraxial shifts to mode frequencies
distort the typical arithmetic spacing of transverse modes,
complicating mode identification,
while millimeter-wave operating frequencies
preclude direct imaging of the modes.
We instead rely on our understanding
of the post-paraxial effects
described in Sec.~\ref{sec:postparax}.

We collect high-resolution transmission~($S_{21}$) spectra of the cavity
over the entire \SIrange{67}{115}{\GHz} pass-band of the measurement setup,
maximizing coupling to cavity modes
by extending the probes
as far as allowed by the translation stages,
giving a distance of~$r = \SI{15.5}{\mm}$
between the probe tips and the cavity axis.
While taking these spectra,
we maintain the cavity at \SI{4}{\K}
in order to keep modes wide relative to our scan resolution
(Sec.~\ref{sec:bcs}).

\begin{figure}
    \centering
    \includegraphics[width=\textwidth]{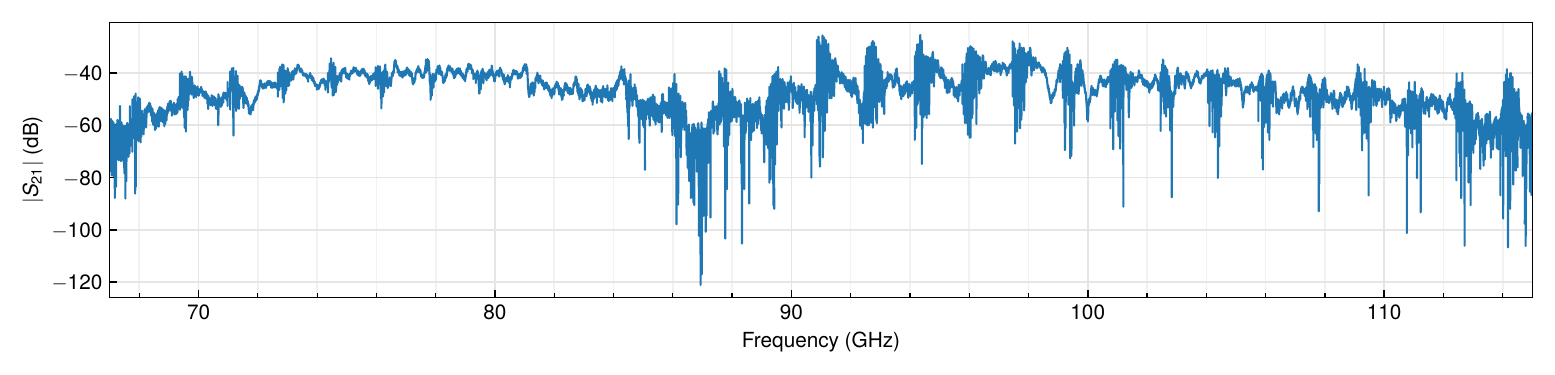}
    \caption{The full $S_{21}$ spectrum of the \lesscfcl{}-geometry
    at \SI{4}{\K} with probes extended as far as possible
    (tips at \SI{15.3}{\mm} from the optical axis).
    The regularly spaced features are groups of Gaussian modes
    spaced by half of the cavity free spectral range (FSR),
    as expected of a near-confocal ($\mean{g} \approx 0$) geometry.}
    \label{fig:full-wide-spectrum}
\end{figure}

\begin{figure}[t]
    \centering
    \includegraphics[width=\textwidth]{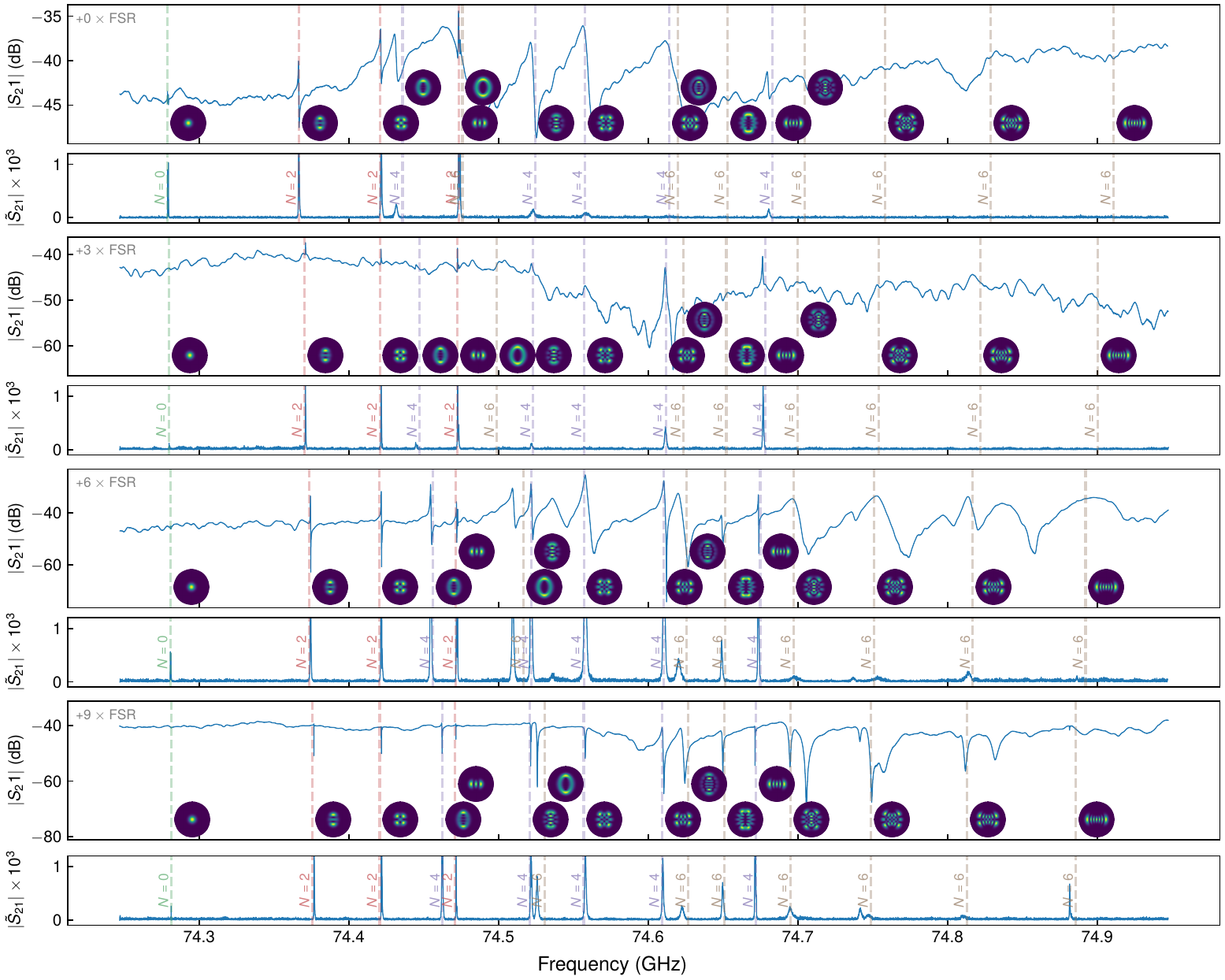}
    \caption{Closeups of some even mode groups
    in the spectrum of the \lesscfcl{}-geometry
    (shown in full in Fig.~\ref{fig:full-wide-spectrum}).
    In this plot, pairs of rows show the same mode group,
    with the first giving the raw spectrum
    and the second a filtered version
    to aid the eye in identifying narrow modes,
    as in Fig.~\ref{fig:cassia-avoided-crossing}\ref{it:raw-data}\==\ref{it:high-passed}.
    The mode groups are shifted by integer multiples of the cavity FSR (\SI{3.2985}{\GHz}) and aligned
    in order to reveal the periodicity of the mode structure.
    We label the modes by their transverse order (up to $N=6$)
    and show in adjacent insets the mode profiles
    inferred from the diagonalization process
    discussed in the text,
    as in Fig.~\ref{fig:cassia-avoided-crossing}.
    Slight errors in the locations of higher-order modes
    are likely due to neglected higher-order corrections in the Hamiltonian.}
    \label{fig:fsr-alignments}
\end{figure}

Since our cavity geometries are near-confocal,
the cavity spectrum consists of groups of modes spaced at $\frac{1}{2}\omega_\text{FSR}$,
where $\omega_\text{FSR} = c/2L$ is the free spectral range (FSR)~%
\cite{siegman1986lasers}.
These groups are readily identified visually
when the probes are well-coupled to the modes
(Fig.~\ref{fig:full-wide-spectrum}).
Each mode group contains only transverse modes of the same parity;
we focus on the even modes (containing $\tem_{00}$),
which have frequencies of around $(n+\frac{1}{2})\omega_\text{FSR}$
for integer~$n$.
Plotting successive even mode groups together
with an offset chosen to approximate the FSR
reveals homologous modes of different longitudinal order,
which we identify as Gaussian modes
(Fig.~\ref{fig:fsr-alignments}).
To make the modes more prominent to the eye,
we take advantage of their high finesse
and apply a high-pass filter to the $S_{21}$ data
to reject the slowly varying background,
which comes from direct crosstalk between the probes.

The $\tem_{00}$ modes are then identified by a few characteristic properties:
\begin{itemize}
    \item
    For the near-confocal geometries considered in this work,
    where $\mean{g} < 0$,
    paraxial theory predicts that
    $\tem_{00}$ modes are lowest in frequency within mode groups.
    This is only approximately true
    after accounting for post-paraxial shifts,
    since higher-order modes can be shifted below the $\tem_{00}$ mode
    when the paraxial transverse mode spacing is too small,
    as is the case in the $\nearcfcl{} = \nearcfclg$ geometry
    closest to confocality
    [Fig.~\ref{fig:cassia-avoided-crossing}\ref{it:anticrossing}].
    \item
    Errant higher-order modes shifted close to $\tem_{00}$ modes
    may be distinguished by their stronger post-paraxial shifts.
    As these shifts are frequency-dependent ($\sim\omega\inv$),
    higher-order modes with negative post-paraxial shifts,
    such as those in near-confocal $\mean{g} < 0$ geometries
    that may be close to $\tem_{00}$ modes,
    will appear to have a slightly higher effective FSR.
    \item
    The $\tem_{00}$ modes are best localized,
    and thus have narrower linewidths
    than higher-order modes.
\end{itemize}


We locate candidate $\tem_{00}$ modes across up to 15 FSRs of data,
then fit a cavity geometry to their position
via Eq.~\ref{eq:post-parax-00}.
We confirm our mode identification
by using this cavity geometry
to predict the frequencies of the higher-order modes in the spectrum,
labeling the entire spectrum in the process
(Fig.~\ref{fig:fsr-alignments}).

\afterpage{\clearpage}

\subsection{Mode mixture reconstruction}
\label{sec:mix-reconstruct}

We reconstruct the mode mixture fractions
in Figs.~\ref{fig:cassia-avoided-crossing}\ref{it:anticrossing}\==\ref{it:anticrossing-coupling}
by fitting a cavity geometry to the observed mode frequencies of the modes involved in the avoided crossing feature
using the theory described in Sec.~\ref{sec:postparax}
together with the following simple model of the hybridization.
For each FSR,
we treat the unmixed modes as eigenvectors~$\ket{\pm}$
of some two-dimensional Hermitian operator
with the frequencies~$\omega_\pm = \omega_0 \pm \delta$ (predicted by the post-paraxial theory) as eigenvalues.
We incorporate coupling between the modes
as an additional off-diagonal constant term~$V$ in the operator
\begin{equation}
    H = \left[\begin{array}{cc}
      \omega_0 + \delta & V \\
      V^* & \omega_0 - \delta
    \end{array}\right].
\end{equation}
Defining $\beta = \tan\inv\frac{\delta}{\abs{V}}$,
we obtain eigenvalues~$\omega_{\pm'} = \omega_0 \pm \abs{V} \sec\beta$
corresponding to normalized eigenvectors
\begin{equation}
    \ket{\pm'} = \pm e^{i\angle V} \sqrt{\frac{1 \pm \sin\beta}{2}} \ket{+} + \sqrt{\frac{1 \mp \sin\beta}{2}} \ket{-},
\end{equation}
so that the power fractions of the coupled eigenmodes are simply $\frac{1\pm\sin\beta}{2}$.


\section{Measurement theory}
\label{sec:measurementtheory}

\subsection{Frequency domain measurements}
\label{sec:vec-fit-and-residues}

In the main text,
we use frequency domain $S_{21}$ measurements
taken with a vector network analyzer (VNA)
to fit mode frequencies (Fig.~\ref{fig:cassia-avoided-crossing}),
determine mode finesses [Fig.~\ref{fig:cassia-avoided-crossing}\ref{it:finesse-dip}],
and extract probe couplings [Figs.~\ref{fig:cassia-avoided-crossing}\ref{it:anticrossing-coupling} and \ref{fig:coupling-strength}\ref{it:coupling}].
We outline our procedure for doing so
in this section.

Frequency domain measurements~$S_{21}(\omega)$
may be viewed as measurements of a Laplace-domain transfer function~$H(s)$
at points~$s = i\omega$.
Modeling cavity resonances as complex pole pairs,
we fit measurements around a cavity resonance
as a superposition of partial fractions
\begin{equation}
    \label{eq:rational-approx}
    H(s) = C + \sum_n \frac{a_n}{s - p_n}
\end{equation}
parametrized by constant~$C$,
residues~$a_n$, and poles~$p_n$,
all complex.
We use the vector fitting algorithm~\cite{gustavsen1999rational}
as implemented by the open-source \texttt{scikit-rf} Python package~\cite{arsenovic2022scikit}
to find optimal parameters,
then feed them as initial conditions
into a nonlinear least-squares routine
to quantify their uncertainties.

\begin{figure}[t]
    \centering
    \includegraphics[width=0.5\linewidth]{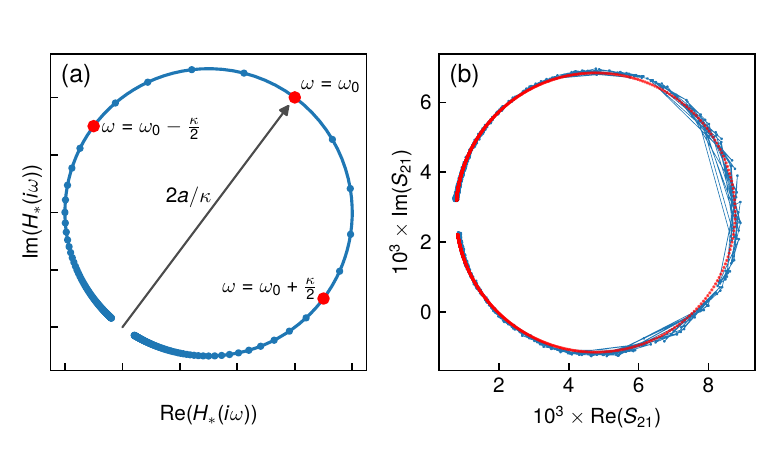}
    \caption[]{%
    \begin{captionenum}
        \item
        \label{it:q-circle-schematic}%
        Schematic of the frequency response~$H_*(i\omega)$ in the complex plane
        due to a transfer function~$H_*(s) = a/(s - p)$
        consisting of a single complex pole~$p = -\frac{1}{2}\kappa + i\omega_0$.
        Points on the curve are marked every $\Delta\omega = \frac{\kappa}{10}$,
        with the center of the resonance
        and the two points at detuning~$\pm\frac{\kappa}{2}$
        highlighted in red.
        \item\label{it:q-circle-data}%
        A fit to the frequency domain trace
        of the $\tem_{21,0,0,x}$ mode
        at $\omega_0 = 2\pi\times\SI{70.981}{\GHz}$
        in the \lesscfcl{}-geometry,
        measured with probes \SI{15.8}{\mm}
        from the cavity axis.
        Both data (blue) and model (red)
        are shown with \SI{50}{\Hz} frequency resolution.
        The fit yields a linewidth of $\kappa = 2\pi\times \SI{7.214(20)}{\kHz}$.
        Cavity vibrations
        cause fast variations in the resonance frequency~%
        $\Delta\omega_0\sim\kappa$,
        visible as the apparent jumping of the data
        between distant points on the circle.
        This distortion slightly biases the resulting fit parameters.
        (See also Fig.~\ref{fig:ringdowns}\ref{it:sweep-vibes}
        for a more extreme example.)
    \end{captionenum}}
    \label{fig:q-circles}
\end{figure}

We fit with more partial fractions
in Eq.~\ref{eq:rational-approx}
than exist cavity modes in our data.
One summand~$H_*(s) = a_*/(s - p_*)$ fits the resonance feature
while the others approximate the nonzero background
(crosstalk between the probes).
The complex pole~$p_*$ in the resonance term~$H_*$
gives the cavity decay rate~$\kappa$
and resonance frequency~$\omega_0$
via its real and imaginary parts:
\begin{equation}
    \label{eq:pole}
    p_* = -\frac{\kappa}{2} + i\omega_0.
\end{equation}
Each resonance term~$H_*(i\omega)$
traces a circle in the complex plane
(the ``$Q$-circle'')
with diameter set by the residue~$a$
via $H_*(\pm i\infty) = 0$ and $H_*(i\omega_0) = 2a_*/\kappa$.
We show a schematic of this frequency response in Fig.~\ref{fig:q-circles},
together with an example fit
to $S_{21}$ data collected in the cavity.

The measured linewidth represents the
\emph{loaded} cavity decay rate~%
$\kappa = \kappa_0 + \kappa_1 + \kappa_2$,
a sum of the intrinsic losses~$\kappa_0$
and losses~$\kappa_{1,2}$ through the probes.
Estimating these external losses
and their dependence on probe position
gives a measure of cavity mode energy distributions
[Figs.~\ref{fig:cassia-avoided-crossing}\ref{it:anticrossing-coupling} and \ref{fig:coupling-strength}\ref{it:coupling}]
via Eq.~\ref{eq:ext-loss}.
This information allows us to ensure the cavity remains undercoupled
($\kappa_{1,2} \ll \kappa_0$)
so that finesse measurements
are not depressed by extrinsic losses.

We constrain the external losses~$\kappa_{1,2}$ using the residue~$a_*$.
Equation~28 of Ref.~\onlinecite{leong2002precise}
gives the diameter of the $S_{21}$ $Q$-circle as
\begin{equation}
    \frac{2\abs{a_*}}{\kappa} = d_{21}
    = \frac{2\sqrt{\kappa_1 \kappa_2}}{\kappa_0 + \kappa_1 + \kappa_2}
\end{equation}
in a system with no insertion loss.
Insertion loss in the probe lines reduces the diameter~$d_{21}$ proportionally,
so we must characterize the insertion loss in our probe lines
to compensate for their effect.
We use a frequency-dependent insertion loss estimate
based on an $S_{21}$ measurement
with the probe lines directly connected to each other
(a ``thru calibration'')
taken at the same cryogenic temperatures
at which we characterize the cavity.
Together with estimates of insertion loss
in components that could not be included in the thru measurement,
we find a frequency-dependent insertion loss
that ranges between \SIlist[list-pair-separator={ and }, list-units=single]{10;25}{\dB}.
(The setup used to characterize the \nearcfcl{}-geometry
differs slightly from Fig.~\ref{fig:block-diagram},
missing the low-noise amplifier
and roughly \SI{50}{\cm} of stainless steel waveguide.
For this variant apparatus,
we estimated an insertion loss of \SI{37.5(2.5)}{\dB}.)

With only transmission measurements,
it is not possible to characterize
the individual couplings~$\kappa_{1,2}$,
so we quote the geometric mean~$\sqrt{\kappa_1 \kappa_2}$
as a measure of the individual probe losses
which should be approximately equal given
given our diametrically positioned probes
and \SI{180}{\degree}-symmetric cavity.
Indeed, the geometric mean lower-bounds the total external losses
by the inequality
\begin{equation}
    \kappa_1 + \kappa_2 \geq 2\sqrt{\kappa_1 \kappa_2}
    =
    2
    \abs{a_*},
\end{equation}
with equality iff $\kappa_1 = \kappa_2$.
In principle,
$Q$-circle diameters obtained from reflection~($S_{11}$) measurements
would give an independent measure of $\kappa_1$,
but these are difficult to measure and calibrate in our cryogenic apparatus.

Frequency domain measurements
are sensitive to residual vibrations in the dilution refrigerator.
In the data of Fig.~\ref{fig:q-circles}\ref{it:q-circle-data},
variations in the mode frequency~$\omega_0$
on the same order as the linewidth~$\kappa$
distort a nominally circular lineshape.
Vibrations induce cavity length variations~$\Delta L \sim \SI{1}{\nm}$
that effectively broaden lineshapes
and make fits to $S_{21}$ data
unsuitable for finesse measurements
when $\lambda/2F \sim \Delta L$,
or $F \gtrsim \num{e+6}$.
For this reason,
we instead use ringdown measurements (Sec.~\ref{sec:ringdown})
for the finesse values quoted in Figs.~\ref{fig:coupling-strength}
and \ref{fig:cavity-perf}.
Empirically, the inferred values of residues~$a_*$
are less affected by vibrations,
though they still suffer from large ($\sim \SI{30}{\percent}$) systematic uncertainties.

\subsection{Probe coupling and local electric field}
\label{sec:probe-coupling}

In the main text, we use the probe couplings~$\kappa_{1,2}$
as a direct measure of the cavity field
and use finite-element simulations of cavity mode fields
to estimate expected probe coupling [Fig.~\ref{fig:coupling-strength}\ref{it:coupling}].
Here we derive the relation between coupling and fields,
following Ref.~\cite{haebel1996couplers}.


We model the coupling to the two probes
as time-harmonic current sources~$I_{1,2}$,
each driving a transmission line of characteristic impedance~$Z = \SI{50}{\ohm}$.
The current comes from the displacement current of the mode field~$\mathbf{E}$
into the bounding surface of the exposed probe tip
\begin{equation}
    I_{1,2} = i\omega \epsilon_0 \iint_{1,2} \mb E \cdot d\mb A,
\end{equation}
and the power loss through the transmission line is simply
\begin{equation}
    P_{1,2} = \frac{1}{2} \lvert I_{1,2}^2 \rvert Z. 
\end{equation}
We obtain the coupling rate by normalizing this power
by the total mode energy~$U$,
which can be calculated from the electric field of the standing wave:
\begin{equation}
    \label{eq:ext-loss}
    \kappa_{1,2} = \frac{P_{1,2}}{U} = \frac{\frac{1}{2} \omega^2 \epsilon_0^2 \abs{\iint_{1,2} \mb E\cdot d\mb A}^2 \abs{Z}}{\iiint \frac{1}{2}\epsilon_0\abs{\mb E}^2 dV}
    =
    \frac{\omega^2 \epsilon_0 \abs{Z} \abs{\iint_{1,2} \mb E\cdot d\mb A}^2}{V_\text{mode} \max\abs{\mb E}^2 }.
\end{equation}
In the last equality,
we introduce the mode volume
\begin{equation}
    V_\text{mode} = \frac{\iiint \abs{\mb E}^2 dV}{\max \abs{\mb E}^2}.
\end{equation}

For a rough coupling estimate,
we may assume that the probe only negligibly perturbs the field strength of the modes
and approximate the displacement current term
by using the electric field of the mode at a single point (the probe tip~$\mb r_{\text{tip}}$).
Since the conducting probe modifies the local field direction,
we take perfectly conducting boundary conditions
over our surface of integration ($\mb E \perp d\mb A$).
Together, these approximations may be summarized as
\begin{equation}
    \Abs{\iint \mb E \cdot d\mb A}^2 \approx \abs{\mb E(\mb r_{\text{tip}})}^2 A^2
\end{equation}
where $A$ is the surface area of the exposed probe tip.
Substituting this expression into Eq.~\ref{eq:ext-loss} yields
\begin{equation}
  \label{eq:probe-coupling}
    \kappa_{1,2} =
    \frac{\omega^2 \epsilon_0 \abs{Z} A^2}{V_\text{mode}  } \frac{\abs{\mb E(\mb r_{\text{tip}, \{1,2\}})}^2}{\max\abs{\mb E}^2}. 
\end{equation}

In Fig.~\ref{fig:coupling-strength}\ref{it:coupling},
we calculate couplings
from the simulated mode fields
under such an approximation.
The surprisingly good quantitative agreement
with our measurements
(e.g.\ couplings that do not dip to zero,
as would occur under this approximation
when $\mb r_\text{tip}$ crosses a node of the mode)
may be understood by the fact that
the finite simulation resolution
introduces a coarse-graining scale
comparable to the probe dimensions.

We use a modified form of Eq.~\ref{eq:probe-coupling} to produce the fit lines
shown in Fig.~\ref{fig:coupling-strength}\ref{it:ringdown-finesse}
as a guide to the eye.
We estimate the cavity loss as $\kappa = \kappa_0 + \kappa_1 + \kappa_2$,
where free parameter~$\kappa_0$ represents intrinsic cavity loss
independent of probe position
and $\kappa_{1,2}$ are dependent on two free parameters $a$~and~$b$ as
\begin{equation}
    \kappa_1 = \kappa_2 = \frac{\omega^2 \epsilon_0 \abs{Z} A^2 }{V_\text{mode}} \frac{a\Abs{\mb E(\frac{r}{b}, z)}^2}{\max \abs{\mb E}^2}
\end{equation}
in order to improve the fit for visual clarity.
Here, we use the paraxial expression for the mode field~$\mathbf{E}$,
unlike in the estimates behind the theory line in
Fig.~\ref{fig:coupling-strength}\ref{it:coupling},
where $\mb E$ comes from the simulation
shown in Fig.~\ref{fig:coupling-strength}\ref{it:q3d-sims}.

\subsection{Ringdown spectroscopy}
\label{sec:ringdown}

\begin{figure}[t]
    \centering
    \includegraphics[width=0.6\linewidth]{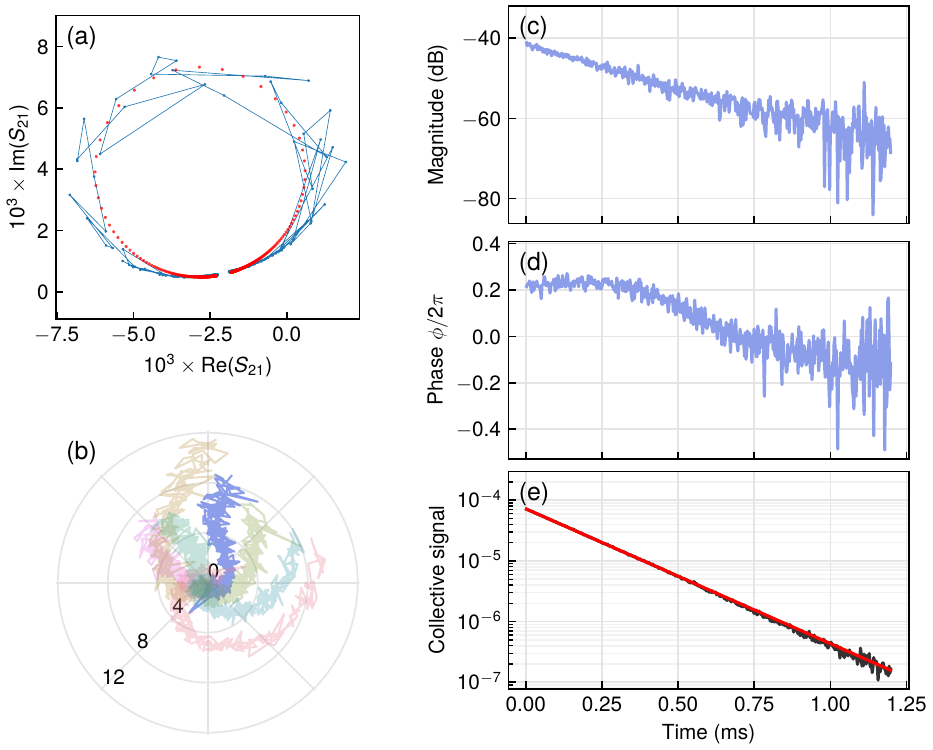}
    \caption[]{%
    Ringdown spectroscopy of the $\tem_{23, 0, 0, x}$ mode
    at \SI{77.579}{\GHz}
    in the \lesscfcl{}-geometry
    with probes at \SI{17.3}{\mm} from the cavity axis.
    \begin{captionenum}
        \item\label{it:sweep-vibes}%
        $S_{21}$ trace of the mode (blue) in the complex plane
        and a fit
        (red, giving linewidth~$\kappa = 2\pi\times \SI{1.768(34)}{\kHz}$),
        both shown with \SI{100}{\Hz} frequency resolution.
        The nominally circular lineshape of the mode
        in the frequency domain suffers significant distortion due to cavity vibrations,
        motivating a vibration-insensitive measurement of mode linewidth.
        \item\label{it:ringdown-ensemble}%
        Repeated ringdowns of the same mode
        plotted in the complex plane.
        The $S_{21}$ values here are multiplied by $10^3$.
        For the shot shown in darker blue,
        we plot
        \item\label{it:rd-db}%
        the magnitude of the signal after subtracting
        the fitted late-time offset~$s_\infty$,
        revealing an exponential decay.
        \item\label{it:rd-phase}%
        The phase~$\phi$
        of the same offset-subtracted signal
        shows vibrations in the cavity length
        through variations in the winding rate.
        \item\label{it:collective}%
        Fitting all \num{200} shots collectively
        as described in the main text
        reveals an exponential decay of the ringdown power
        and a linewidth~$\kappa = 2\pi\times\SI{813.3(10)}{\Hz}$
        narrower than the frequency-domain measurement by a factor of two.
        This linewidth is shown as a finesse
        in Fig.~\ref{fig:coupling-strength}\ref{it:ringdown-finesse}.
        \end{captionenum}
    }
    \label{fig:ringdowns}
\end{figure}

In order to measure the true finesse of the cavity modes
in the presence of cryostat vibrations
[Fig.~\ref{fig:ringdowns}\ref{it:sweep-vibes}],
we turn to ringdown spectroscopy
[Fig.~\ref{fig:ringdowns}\ref{it:ringdown-ensemble}\==\ref{it:collective}].
To probe a mode of frequency~$\omega(t)$,
which varies in time due to vibrations,
a single-tone drive of frequency~$\omega_0$ on probe~1
pumps the cavity mode with power before it is abruptly cut off,
at which point we begin measuring the leakage signal through probe~2
with the VNA.
The resulting (complex) signal takes the form
\begin{equation}
    \label{eq:single-ringdown-random}
    s(t) = s_\infty + s_0 e^{-\frac{1}{2}\kappa t + i \int_0^t \delta(t') \,dt'} + \tilde\epsilon(t),
\end{equation}
where $s_\infty$ represents some small nonzero signal in our system
due to crosstalk in the probe electronics
and imperfect isolation in the off-state of the switch,
$s_0$~is an initial signal level that varies shot-to-shot,
$\kappa$~is the cavity decay rate,
$\delta(t) = \omega(t) - \omega_0$~%
the instantaneous detuning of the cavity mode from the probe,
and $\tilde\epsilon(t)$~for each time~$t$ is
an independent identically distributed,
circularly-symmetric
complex Gaussian random variable
with mean zero and
variance~$\langle\abs{\tilde\epsilon}^2\rangle = \sigma^2$.
We plot several such traces for a single mode
in Fig.~\ref{fig:ringdowns}\ref{it:ringdown-ensemble},
which shows a nonzero late-time value~$s_\infty$
and the shot-to-shot variations in initial amplitude~$s_0$.
These traces often take the form of spirals;
indeed, in the special case of constant detuning~$\delta(t) = \delta$,
Eq.~\ref{eq:single-ringdown-random} traces a logarithmic spiral in the complex plane.

To extract the decay rate~$\kappa$,
we remove the dependence on $\delta(t)$
by noting that in expectation,
\begin{equation}
    \label{eq:single-ringdown}
    \bigl\langle \Abs{s(t) - s_\infty}^2 \bigr\rangle = \abs{s_0}^2 e^{-\kappa t} + \sigma^2.
\end{equation}
For an appropriately chosen offset~$s_\infty$,
$\abs{s(t) - s_\infty}^2$ should fit well
to a decaying exponential
plus a positive constant~$\sigma^2$
representing the noise power.
We therefore analyze ringdown curves
by fitting $\abs{s(t) - s_\infty}^2$
with the model of Eq.~\ref{eq:single-ringdown}
while varying the offset~$s_\infty$ to optimize the fit.
Fitting one of the shots
in Fig.~\ref{fig:ringdowns}\ref{it:ringdown-ensemble}
in this manner,
we show the exponential decay in the amplitude
of the offset-subtracted signal~$s(t) - s_\infty$
in Fig.~\ref{fig:ringdowns}\ref{it:rd-db}.

Figure~\ref{fig:ringdowns}\ref{it:rd-phase}
plots the phase of the same signal,
revealing the amplitude of vibrations in the cavity.
Indeed, Eq.~\ref{eq:single-ringdown-random}
gives the instantaneous mode--probe detuning~$\delta(t)$ as
\begin{equation}
    \delta(t) = \frac{d}{dt}\angle \bigl(s(t) - s_\infty\bigr)
\end{equation}
in the absence of noise.
The slight increase in phase at early times~$t < \SI{0.2}{\ms}$
gives $\delta \approx 2\pi\times \SI{200}{\Hz}$,
while the decrease between \SIrange{0.3}{0.6}{\ms}
gives a detuning of \SI{-500}{\Hz},
for a frequency variation of~$\Delta\omega \approx 2\pi\times\SI{700}{\Hz}$
and a length variation~$\Delta L = L \frac{\Delta\omega}{\omega} \approx \SI{0.4}{\nm}$.


\newcommand{\ringdownind}{j}

For each ringdown measurement,
we average $N \sim \numrange{e2}{e3}$ shots $s^{(\ringdownind)}(t)$
as follows.
We subtract a common value of $s_\infty$ from each,
then fit the ensemble average of $\abs{s^{(\ringdownind)}(t) - s_\infty}^2$
to the form of Eq.~\ref{eq:single-ringdown},
varying $s_\infty$ to obtain the best fit
as before.
The prefactor on the exponential is then an estimator
for the expectation value $\langle \abs{s_0^{(\ringdownind)}}^2\rangle$.
Importantly,
we must average traces only after removing phase information:
the complex traces~$s^{(\ringdownind)}(t)$ are not mutually phase coherent
due to shot-to-shot variation in the mode frequency
from cavity vibrations,
so averaging the $s^{(\ringdownind)}(t)$ directly
would produce a trace with faster decay.
Figure~\ref{fig:ringdowns}\ref{it:collective}
illustrates the reduced noise in ensemble-averaged data
(shown with the noise power~$\sigma^2$ subtracted)
compared to single-shot data [Figs.~\ref{fig:ringdowns}\ref{it:rd-db} and \ref{it:rd-phase}].
The ringdown curve shown in Fig.~\ref{fig:cavity-perf}\ref{it:ringdown-inset}
is precisely such an ensemble-averaged noise-subtracted signal,
normalized as a fractional power.

\section{Finite-element simulations}
\label{sec:sims}

This section provides additional technical details on the finite-element simulations mentioned in the main text.
These simulations
supported measurements of probe--mode coupling [Fig.~\ref{fig:coupling-strength}\ref{it:coupling}],
which revealed improvement in mode localization
upon detuning the cavity geometry away from confocality.
Mode field patterns obtained from the simulations
illustrated the qualitative contrast in spatial structure
between modes in the \nearcfcl{}- and \lesscfcl{}-geometries [Fig.~\ref{fig:coupling-strength}\ref{it:q3d-sims}]
and yielded predictions of coupling strength
in excellent quantitative agreement
with measurements in the \lesscfcl{}-geometry.

The superconducting cavity presents two challenges
to robust eigenmode simulation:
large simulation volume
and high~$Q$~factors.
Direct simulation of a volume~$V$
large compared to the mode wavelength~$\lambda$
($V \gg \lambda^3$, an ``overmoded'' geometry)
is computationally expensive.
High~$Q$~factors, meanwhile, make estimating mode losses~$\kappa$
numerically challenging.
Eigenmode simulations give mode loss
by computing complex mode frequencies~%
$\tilde\omega = \omega + \frac{1}{2}i\kappa = \omega(1 + \frac{i}{2Q})$,
where $\omega$~is the (real) mode frequency
(cf.\ Eq.~\ref{eq:pole}).
In the limit~$Q\gg 1$,
accurately determining the loss~$\kappa$
then requires finding the complex frequency~$\tilde\omega$
to high fractional accuracy.
Reliable estimation of low mode losses is important
given our appeal to simulations
to contrast the amount of leakage power
expected in the \nearcfcl{}- and \lesscfcl{}-geometries
[Fig.~\ref{fig:coupling-strength}\ref{it:q3d-sims}].


To address the difficulties posed
by the overmoded cavity geometry
and the high cavity $Q$ factors,
we run beyond-quadratic-order (``high-order'')
quasi-3D finite element simulations.
Here, ``order'' refers both to \emph{mesh order},
the polynomial degree of the mesh elements
used to discretize the simulation geometry,
and to \emph{discretization order},
the degree of the shape functions---%
polynomials chosen as basis functions
for approximating the electric field of an eigenmode
within each mesh element.
High-order simulations have previously been used
in a modal analysis of the similar high-finesse cavity of Ref.~\cite{kuhr2007ultrahigh},
which found that both high mesh order and high discretization order was necessary
for accurate simulation of the low mode losses~\cite{marsic2018modal}.
Furthermore, 
our simulations approximate the cavity
as cylindrically symmetric,
replacing the toroidal mirrors with spherical mirrors
matching the harmonic mean radius of curvature~$\mean{R}$
of the toroids.
This symmetrization permits the use of ``quasi-3D'' methods
that reduce the simulation problem to two dimensions
to avoid the computational cost
of simulating the overmoded cavity geometry~\cite{schnaubelt2021comparison}.

Our simulations use the high-order simulation code
of the \texttt{small\_fem} library~\cite{marsic2018modal,schnaubelt2021comparison}.
Each simulation considers the cavity centered in a cylindrical simulation volume
extending~$3\lambda$ past the outer edge of the mirrors
in order to ensure that the simulated mode
is not perturbed by the simulation boundary.
Of this padding, the outer~$1\lambda$ is a perfectly matched layer (PML)
that absorbs incident radiation,
representing the open boundary conditions of the cavity.
The additional inner~$2\lambda$ is a precaution against
residual reflections from the vacuum--PML interface
and from the simulation boundary,
which arise from discretization
and from the finite thickness of the PML.
We mesh the geometry at fourth order
with a mesh length of~$\lambda/8$,
using fourth-order basis functions
and the quasi-3D transformation~``TA'' of Ref.~\cite{schnaubelt2021comparison}.
The resulting simulations provide fast, reliable estimates
of mode losses,
raising the possibility of
numerical optimization of mirror profiles
for further suppression of mode mixing
in future work.
Indeed, runtimes of a few minutes
allow simulations to be incorporated
into numerical optimization algorithms
as calls to a subroutine
that quickly calculates the mode loss
of any hypothetical cavity geometry.

The simulations
qualitatively illustrate the effects of mode mixing
that we observe near confocality,
but can understate the degree of mode mixing
present in our measurements
due to the simplifying assumption of rotational symmetry.
Specifically, for the \nearcfcl-geometry mode
shown in Fig.~\ref{fig:coupling-strength}\ref{it:q3d-sims},
the dominant contribution to mode hybridization
is the higher-order mode
in the avoided crossing of Fig.~\ref{fig:cassia-avoided-crossing},
which involves modes of differing angular momentum
that would not hybridize in a rotationally symmetric system.
$\tem_{00}$ modes have orbital angular momentum~$\ell = 0$;
additionally including photon spin~$s = \pm 1$
(manifested as the polarization of the mode)
gives a total angular momentum~$J = \pm 1$.
The fitting procedure
described in Sec.~\ref{sec:mode-id},
meanwhile,
shows that the intruding higher-order modes
predominantly consist of the Laguerre-Gauss modes~$\tem_{p=0,\ell=\pm4,s=\pm1}$
of radial index~$p=0$,
orbital angular momentum~$\ell=\pm4$,
and circular polarization~$s=\pm1$
with the same chirality as $\ell$.
In the symmetrized geometry,
this series of modes becomes exactly $\tem_{0,\pm4,\pm1}$;
the resulting $J = \pm 5$ modes
therefore cannot couple with $\tem_{00}$ modes.
Nonetheless, we observe increased leakage power in the \nearcfcl{}-simulation
versus the \lesscfcl{}-simulations
due to the smaller frequency gaps to modes with symmetry-allowed couplings.

\section{Cavity loss estimates}
\label{sec:losses}


This section discusses in greater detail
a number of loss mechanisms in the superconducting cavity
that are mentioned in the main text.
We begin with some fundamentals in Sec.~\ref{sec:this-is-loss},
then turn to simple estimates of diffractive losses
in Sec.~\ref{sec:diffractive-loss}.
We then consider two superconductor loss channels
that drive loss inside the niobium layer of the cavity mirrors:
a temperature-dependent surface resistance~$R_\text{BCS}(T)$
described by
BCS theory (Sec.~\ref{sec:bcs})
and a loss~$R_\text{flux}$
arising from trapped magnetic flux
(Sec.~\ref{sec:vortices}).
We only sketch the relevant theory for these superconductor loss channels below,
directing the interested reader
to the excellent review of Ref.~\onlinecite{gurevich2017theory}
for a fuller exposition.

\subsection{Loss in Fabry--P\'erot cavities}
\label{sec:this-is-loss}


In a symmetric Fabry--P\'erot cavity,
each loss mechanism~$j$ can be quantified
by the fraction~$T_j$ of incident power it dissipates
upon reflection off of one cavity mirror.
The effective total loss fraction is simply the sum~$T = \sum_j T_j$
when losses are weak~($T_j \ll 1$),
true for all loss mechanisms considered in this work.
Finesse~$F$ is related to total one-way loss as $F = \pi/T$.
The individual loss mechanisms
may also be quantified with the cavity finesse~$F_j = \pi/T_j$
attainable if other loss mechanisms were absent,
and are related to the overall finesse by $F\inv = \sum_j F_j\inv$.

Following common practice in the superconducting rf literature,
we present superconductor losses in terms of surface resistance~$R_s$,
the real part of surface impedance~$Z_s$.
Surface resistance may straightforwardly be related to finesse
by considering the classic problem
of wave propagation at an impedance boundary.
Indeed,
for radiation with wavelength
much smaller than the mirror curvature,
reflection off the cavity mirrors
can be modeled as plane wave radiation propagating in vacuum
(with wave impedance~$Z_0 = \mu_0 c = \SI{377}{\ohm}$)
incident on a sheet of material of impedance~$Z_s$.
If the incident wave
has electric and magnetic field amplitudes~%
$E_0$ and $H_0 = E_0 / Z_0$,
the resulting transmitted wave
has approximately double the magnetic field amplitude,
since the complex transmission coefficient is
\begin{equation}
    t = \frac{2Z_0}{Z_0 + Z_s} \approx 2
\end{equation}
in the limit~$\abs{Z_s} \ll Z_0$
of low losses.

We are interested in comparing
the intensities of the incident and transmitted waves.
Intensity of a plane wave
with complex field amplitudes~$E$ and $H$
in a medium of wave impedance~$Z$
is given by the magnitude of the time-averaged Poynting vector~%
$\langle S \rangle = \frac{1}{2} \Re(EH^*) = \frac{1}{2} \Re(Z) \abs{H}^2$.
The transmitted wave in our problem
thus corresponds to a fractional power loss
\begin{equation}
    T = \frac{\frac{1}{2} \Re(Z_s)\abs{tH_0}^2}{\frac{1}{2} \Re(Z_0)\abs{H_0}^2}
    =
    \frac{4R_s}{Z_0}
\end{equation}
and a finesse
\begin{equation}
    \label{eq:finesse-surfaceresistance}
    F = \frac{\pi Z_0}{4 R_s}.
\end{equation}

In the superconducting rf literature,
cavity mode losses are usually quoted as $Q$ factors
related to surface resistance~$R_s$
by a geometry-dependent factor~$G$
as $Q = G/R_s$.
Equation~\ref{eq:finesse-surfaceresistance}
is thus equivalent to deriving a geometry factor
\begin{equation}
    G = QR_s = F \frac{\omega}{\omega_\text{FSR}} R_s
    =
    \frac{1}{4} k L Z_0
\end{equation}
for Fabry--P\'erot cavities
of length~$L$,
where $\omega$ and $\omega_\text{FSR}$
are respectively the mode frequency and free spectral range,
consistent with previous work \cite{klein1987proposal}.

\subsection{Diffractive losses}
\label{sec:diffractive-loss}

We now give a rough treatment
of the diffractive loss in the cavity,
which arises both from clipping of the cavity mode
on mirrors of finite transverse extent
and from the microscopic roughness of the mirror surfaces.
Our estimates of clipping loss guided the cavity design
[Fig.~\ref{fig:cavity-geometry}\ref{it:design-space}],
while the result for diffraction from surface roughness
sets the limit on the maximum finesse achievable with our cavity mirrors
in Fig.~\ref{fig:cavity-perf}.

Consider a simple model of clipping loss.
We assume a pure Gaussian beam incident on a mirror of transverse radius~$r$
with a spot size~$w$ at the mirror edge,
defined as the radius at which intensity falls off to $1/e^2$ of the (maximum) value on-axis.
Integrating the Gaussian intensity profile
outside the radius~$r$
yields
a finesse limit of
\begin{equation}
    \label{eq:clipping-finesse}
    F_\text{clip} = \pi e^{2r^2/w^2}.
\end{equation}
We use this result in Fig.~\ref{fig:cavity-geometry}\ref{it:design-space};
for each cavity geometry in our design space,
we show the transverse optical access available for imaging
given mirrors sized such that clipping loss (estimated as above)
limits the cavity to a finesse~$F_\text{clip} = \num{e10}$.
The simple analytic estimate of Eq.~\ref{eq:clipping-finesse}
is sufficient to guide our cavity design,
though we note that the effect of finite mirror size
is more properly treated as an operator coupling
various cavity modes,
as considered in Ref.~\cite{kleckner2010diffraction}.

An extended version of Eq.~\ref{eq:clipping-finesse}
yields the fit lines
shown in Fig.~\ref{fig:cavity-perf}\ref{it:best-fins},
from which we extract the plateau value of the cavity finesse
at high frequencies.
There, we estimate overall finesse as a combination of a limiting finesse~$F_{\lim}$
representing losses due to other effects
and a modified clipping finesse
\begin{equation}
    F_{\text{clip}}^*(b) = \pi e^{2r^2/(bw)^2}.
\end{equation}
Here, $F_{\lim}$ is a free parameter,
as is the factor~$b$
representing an effective expansion of the mode
in order to qualitatively capture the effects of mode mixing.
The fits in Fig.~\ref{fig:cavity-perf}\ref{it:best-fins}
give plateau values~$F_{\lim, \{1,2\}} = \numlist[list-exponents=combine-bracket]{6.16(21)e+7;5.74(10)e+7}$
for the \lesscfcl{}- and \hadriana{}-geometries,
respectively.
The inverse variance--weighted average of these values
produces the concordance value of \maxfinesse{}
we quote in the main text.

We may also estimate diffractive loss
due to diffuse scattering off of the mirror surface,
deriving the expression used to estimate
the roughness-limited performance of the cavity
in Fig.~\ref{fig:cavity-perf}\ref{it:best-fins}.
A flat surface with rms surface roughness~$h_\text{rms}$
reflects a fraction~$T_\text{surf} = (2kh_\text{rms})^2$
of incident plane waves of wavenumber~$k$~\cite{siegman1986lasers,winkler1994light},
immediately yielding the expression
\begin{equation}
    F_\text{surf} = \frac{\pi}{4k^2 h_{\text{rms}}^2}
\end{equation}
given in the main text.
This expression holds approximately
for our curved cavity mirrors,
which are roughly planar at the millimeter scale
of the radiation wavelengths~$\lambda$.

\subsection{Bardeen--Cooper--Schrieffer resistance}
\label{sec:bcs}

In this section,
we present a brief review of
BCS  
resistance~$R_\text{BCS}$,
which describes dissipation of rf radiation
incident on a superconductor
due to the finite population of
thermally dissociated Cooper pairs
(quasiparticles)
at nonzero temperature,
loosely following Refs.~\cite{schmueser2006basic,gurevich2017theory}.
Our understanding of BCS resistance
explains the temperature scaling
of the cavity finesse
at temperatures~$T \gtrsim \SI{1.5}{\K}$
[Fig.~\ref{fig:cavity-perf}\ref{it:temp-dep}].

When treating the electromagnetic response of superconductors,
two length scales are relevant:
the coherence length~$\xi$
describing the delocalization of charge carriers
and the penetration depth~$\london$
over which incident fields at depth~$z$
are exponentially attenuated as $e^{-z/\london}$,
as predicted by the London equations.
These length scales are material properties;
in pure niobium, the coherence length is~$\xi_0 \approx \SI{40}{\nm}$
while the penetration depth is~$\london_0 \approx \SI{40}{\nm}$
at absolute zero.
The presence of impurities affects both,
decreasing coherence length~$\xi$
and increasing penetration depth~$\london$.
Nonzero temperature~$T$ also increases~$\london$.
We shall consider the commonly assumed
``local London limit''~$\london \gg \xi$,
where the penetration depth is much longer
than the coherence length.
Though we will find this assumption not entirely self-consistent,
we nonetheless obtain a reasonable model
for the temperature dependence of the cavity finesse
shown in Fig.~\ref{fig:cavity-perf}\ref{it:temp-dep}.

In the local London limit,
the response of the cavity mirrors
may be characterized by a complex, frequency-dependent conductivity~$\sigma$:
fields do not vary appreciably over the length scale of a delocalized charge carrier
such that we recover the familiar ohmic relation~$\mb J = \sigma \mb E$
between current density~$\mb J$
and electric field~$\mb E$,
both of which we take to be time-harmonic with frequency~$\omega$.
The complex conductivity~$\sigma$ may be understood using a \emph{two-fluid model},
in which charge carriers in a superconductor
are modeled as a mixture of a normal fluid
and an inviscid superfluid.
These two components,
corresponding respectively to
dissociated and bound Cooper pairs
at a microscopic level,
independently contribute to the conductivity as $\sigma = \sigma_1 - i\sigma_2$.
The normal fluid gives rise to the small dissipative real part~$\sigma_1$
that we will obtain from BCS theory,
while the superfluid produces
a purely imaginary contribution~$-i\sigma_2$,
since its velocity~$\mb v$
is $\pi/2$ out of phase
with the drive field~$\mb E \propto \dot{\mb v} = i\omega \mb v \propto i\mb J$.
We determine~$\sigma_2$ in terms of
the penetration depth~$\london$
by deriving the latter
as an effective skin depth
from the conductivity~$\sigma$:
\begin{equation}
    \label{eq:sigma2}
    \london^2 = -\frac{i}{\mu_0 \omega \sigma} \approx \frac{1}{\mu_0 \omega \sigma_2}.
\end{equation}
The last expression assumes weak ohmic losses~$\sigma_1 \ll \sigma_2$,
a limit easily attained for our temperatures~$T < T_c/2$.
As we shall soon find,
only a small fraction of the charge carriers in the superconductor
populate the normal fluid.

From the complex conductivity~$\sigma$,
classical electrodynamics
gives the surface impedance~$Z_\text{BCS}$ as
\begin{equation}
    Z_\text{BCS} = \sqrt{\frac{i\mu_0 \omega}{\sigma}}
    \approx
    \sqrt{\frac{\mu_0 \omega}{\sigma_2}} \left(\frac{\sigma_1}{2\sigma_2} + i\right),
\end{equation}
where we have found the real and imaginary parts
to leading order in $\sigma_1/\sigma_2$~\cite{jackson2021classical}.
Expressing the superfluid conductivity~$\sigma_2$
in terms of the penetration depth~$\london$ (Eq.~\ref{eq:sigma2}),
we arrive at the formula for the surface resistance
\begin{equation}
  \label{eq:bcs-gurevich}
    R_\text{BCS} = \Re(Z_\text{BCS}) =
    \frac{\mu_0^2 \omega^2 \london^3}{2\rho_n} \frac{\sigma_1}{\sigma_n},
\end{equation}
where $\rho_n$~and~$\sigma_n$
are the normal state resistivity and conductivity of the material.
We have introduced the extra factor of $\rho_n \sigma_n = 1$
in the denominator in order to obtain the conductivity ratio~$\sigma_1/\sigma_n$,
which we shall compute from the microscopic theory.
We fix the normal-state resistivity to~$\rho_n = \SI{2.8}{\nano\ohm\meter}$
based on dc measurements of witness samples prepared during mirror fabrication~(Sec.~\ref{sec:mirror-fab}).

BCS theory gives the conductivity ratio~$\sigma_1/\sigma_n$
as an integral
\begin{equation}
    \label{eq:conductivity-ratio}
    \frac{\sigma_1}{\sigma_n} = \frac{2}{\hbar\omega} \int_\Delta^\infty
        \frac{
          (\epsilon^2 + \Delta^2 + \hbar\omega\epsilon)
          \bigl[f_0(\epsilon) - f_0(\epsilon + \hbar\omega)\bigr]
        }
        {\sqrt{\epsilon^2 - \Delta^2}\sqrt{(\epsilon+\hbar\omega)^2-\Delta^2}} d\epsilon
\end{equation}
over the energy~$\epsilon$ of normal fluid charge carriers,
where $\Delta = k_B \times \SI{17.67}{\K}$
is the superconducting gap in niobium
and
$f_0(\epsilon) = (1 + e^{\epsilon/k_B T})\inv$
is the Fermi--Dirac distribution function.
Equation~\ref{eq:conductivity-ratio} only assumes
that drive frequencies~$\omega$
are below the threshold~$2\Delta/\hbar$
required for the drive to directly dissociate Cooper pairs,
as is true throughout this work:
$2\Delta/h = \SI{700}{\GHz}$ in niobium.

If one assumes further
that temperatures are far below the critical temperature~$T \ll T_c$
and that drive frequencies are low compared to the characteristic thermal energy
($\hbar\omega \ll k_B T$),
the integral of Eq.~\ref{eq:conductivity-ratio}
simplifies considerably
to yield the well-known approximation
\begin{equation}
    \label{eq:bcs-wikipedia}
    R_\text{BCS}(T) \approx \frac{A\omega^2}{T} e^{-\Delta/k_B T},
\end{equation}
where $A$ is a material- and purity-dependent constant
subsuming prefactors from Eq.~\ref{eq:bcs-gurevich}
and from the integral of Eq.~\ref{eq:conductivity-ratio}.
The Boltzmann scaling~$R_\text{BCS} \sim e^{-\Delta/k_B T}$
dominates the temperature dependence of BCS surface resistance,
and may be understood intuitively.
The ratio~$\sigma_1/\sigma_n$
appearing in Eq.~\ref{eq:bcs-gurevich}
should be given roughly by
the fraction of charge carriers in the normal fluid,
or alternatively,
the fraction of Cooper pairs
that are thermally dissociated.
The pair dissociation energy of
$\Delta$~per electron
then produces a Boltzmann scaling~$R_\text{BCS} \sim \sigma_1/\sigma_n \sim e^{-\Delta/k_B T}$.

\begin{figure}
    \centering
    \includegraphics[width=0.45\textwidth]{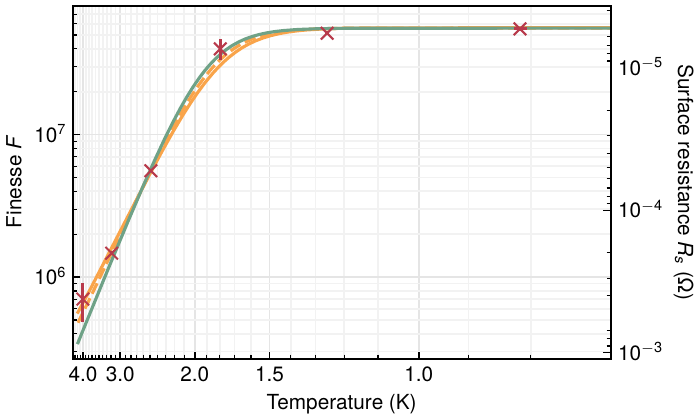}
    \caption{Comparison of methods of computing BCS loss
    for fitting the temperature dependence of mode finesse.
    Arrhenius plot
    (logarithmic $y$-scale
    and reciprocal temperature scale)
    shows Boltzmann scaling~$e^{-\Delta/k_B T}$
    as lines of slope set by energy~$\Delta$.
    We consider finesse measurements
    for the $\tem_{28,0,0,y}$ mode at \SI{94.073}{\GHz}
    in the \hadriana{}-geometry
    [red crosses,
    also shown in Fig.~\ref{fig:cavity-perf}\ref{it:temp-dep}],
    and model the surface resistance
    as a sum $R_0 + R_\mathrm{BCS}(T)$
    of a residual resistance~$R_0$ (a free parameter)
    and the temperature-dependent BCS resistance~$R_\mathrm{BCS}(T)$.
    BCS resistance is estimated
    either with the numerical method described in the text
    and shown in Fig.~\ref{fig:cavity-perf}\ref{it:temp-dep}
    (\BCSNumericColor{})
    or with the well-known approximation
    of Eq.~\ref{eq:bcs-wikipedia}
    (solid \BCSBasicColor{});
    in each case the prefactor on $R_\mathrm{BCS}$
    is a free parameter
    and the superconducting gap~$\Delta$ is
    fixed to the $k_B \times \SI{17.67}{\K}$ value in niobium.
    Further allowing the gap~$\Delta$ in the approximate expression
    to float as a third free parameter
    recovers a stronger scaling with temperature
    by fitting a higher value~$\Delta = k_B \times \SI{18.9(6)}{\K}$
    (dashed \BCSBasicColor{}).
    }
    \label{fig:bcs-exprs}
\end{figure}


In our analysis,
we always compute the conductivity fraction~$\sigma_1/\sigma_n$
by numerically integrating Eq.~\ref{eq:conductivity-ratio},
as Eq.~\ref{eq:bcs-wikipedia} breaks down
at temperatures~$T \gtrsim \SI{1}{\K}$
for millimeter-wave frequencies:
the frequency~$\omega = 2\pi\times\SI{94.073}{\GHz}$
of the mode shown in Fig.~\ref{fig:cavity-perf}\ref{it:temp-dep},
for instance,
yields $\hbar\omega/k_B = \SI{4.5}{\K}$.
The Arrhenius plot of Fig.~\ref{fig:bcs-exprs}
compares the results of fitting the temperature dependence shown in Fig.~\ref{fig:cavity-perf}\ref{it:temp-dep}
using our numerical approach (\BCSNumericColor{})
and using the approximation of Eq.~\ref{eq:bcs-wikipedia}
(\BCSBasicColor{}, with prefactor~$A$ allowed to float).
While both are consistent with the data,
the numerical approach is better motivated physically
and allows us to extract the penetration depth~$\london$
through the prefactor of Eq.~\ref{eq:bcs-gurevich}.
The differences between the two approaches
may become important
in cavities with even lower losses:
the numerical approach predicts a slightly stronger scaling of the cavity loss
with temperature
than the approximation of Eq.~\ref{eq:bcs-wikipedia},
an inconsistency which becomes more evident
in finesse measurements spanning several orders of magnitudes
and which appears to the approximate approach
as an effective increase
in the superconducting gap~$\Delta$
(Fig.~\ref{fig:bcs-exprs},
dashed \BCSBasicColor{}),
as observed in a similar superconducting cavity
in Ref.~\cite{kuhr2007ultrahigh}.


We take care to account for the dependence of penetration depth~$\london$
in Eq.~\ref{eq:bcs-gurevich}
on both temperature and material purity.
First, the penetration depth weakly varies with temperature
following the empirical form~%
$\london(T) = \london_{0} [1 - (T/T_c)^4]^{-1/2}$~%
\cite{tinkham1996introduction}.
Penetration depth further depends on material purity as
\begin{equation}
  \label{eq:lambda-impure}
    \frac{1}{\london^2} = \frac{1}{a\london_\text{pure}^2}
    \left(
      \frac\pi2 - \frac{\cos\inv a}{\sqrt{1-a^2}}
    \right),
\end{equation}
where $\london_\text{pure}$~is the pure material penetration depth
and where parameter~$a = \pi\xi_0/2\ell$
quantifies the material impurity
using the mean free path~$\ell$ of charge carriers,
determined using the measured normal-state resistivity~$\rho_n$
and the Drude formula
\begin{equation}
    \label{eq:drude}
    \rho_n = \frac{p_F}{n_0 e^2 \ell}.
\end{equation}
Here,
$p_F = \hbar \sqrt[3]{3\pi^2 n_0}$~is the Fermi momentum,
$n_0$~the charge carrier density,
and $e$~the elementary charge.

To fit the temperature dependence
of the cavity finesse
[Fig.~\ref{fig:cavity-perf}\ref{it:temp-dep}], then,
we consider the functional form
\begin{equation}
    \label{eq:fig4b-model}
    F = \frac{\pi Z_0}{4\bigl(R_0 + R_\text{BCS}(T; \london_0)\bigr)},
\end{equation}
obtained from Eq.~\ref{eq:finesse-surfaceresistance}
with a surface resistance
consisting of a residual resistance~$R_0$
quantifying temperature-independent losses
and the BCS resistance given by Eq.~\ref{eq:bcs-gurevich}.
We leave $R_0$~and the absolute-zero pure-material penetration depth~$\london_0$
as free parameters.
The residual resistance~$R_0$
sets the height of the finesse plateau,
while the depth~$\london_0$
determines the absolute level
of the finesse curve
below the plateau.
The fit to the temperature dependence
of the cavity finesse in Fig.~\ref{fig:cavity-perf}\ref{it:temp-dep}
yields~$\london_0 = \SI{37}{\nm}$.
From the measured resistivity~$\rho_n$,
we compute a mean free path~$\ell \approx \SI{320}{\nm}$
(Eq.~\ref{eq:drude})
much longer than the pure niobium coherence length~$\xi_0 = \SI{40}{\nm}$.
Therefore, both the penetration depth~$\london$
and coherence length~$\xi$
are close to their pure values:
$\london = \SI{40}{\nm}$ (Eq.~\ref{eq:lambda-impure})
and $\xi\inv = \xi_0\inv + \ell\inv = (\SI{36}{\nm})\inv$.
Though this violates the local London limit~$\xi \ll \london$,
we find that Eq.~\ref{eq:fig4b-model}
adequately models
the temperature scaling of the cavity finesse.

\subsection{Trapped vortex loss}
\label{sec:vortices}

In the main text,
we interpret the cavity loss
at high frequencies~$\omega\gtrsim 2\pi\times\SI{90}{\GHz}$
as arising primarily from trapped Abrikosov vortices
[Fig.~\ref{fig:cavity-perf}\ref{it:best-fins}];
here, we describe and estimate this vortex loss,
following Ref.~\cite{gurevich2017theory}.

When type-II superconductors
are cooled through the superconducting transition temperature~$T_c$
under an external magnetic field,
vortices of supercurrent form
surrounding single quanta of trapped magnetic flux.
These vortices are thermodynamically disfavored below~$T_c$,
but can be pinned in place by impurities.
Each vortex surrounds a tube of normal-state material,
and these normal-state tubes collectively produce
the excess loss that we seek to estimate.
The resulting surface resistance~$R_\text{flux}$
is thus proportional
both to the resistivity~$\rho_n$ of these normal-state tubes
and to the trapped magnetic field~$B_\perp$
normal to the mirrors,
which determines the vortex density.
We measure the normal-state resistivity~$\rho_n$
using witness samples prepared during mirror fabrication
(Sec.~\ref{sec:mirror-fab}).
Magnetometer measurements
adjacent to the dilution refrigerator at room temperature,
meanwhile,
yield an estimate~$B_\perp = \SI{0.22(3)}{\gauss}$.

\begin{figure}
    \centering
    \includegraphics[width=0.5\textwidth]{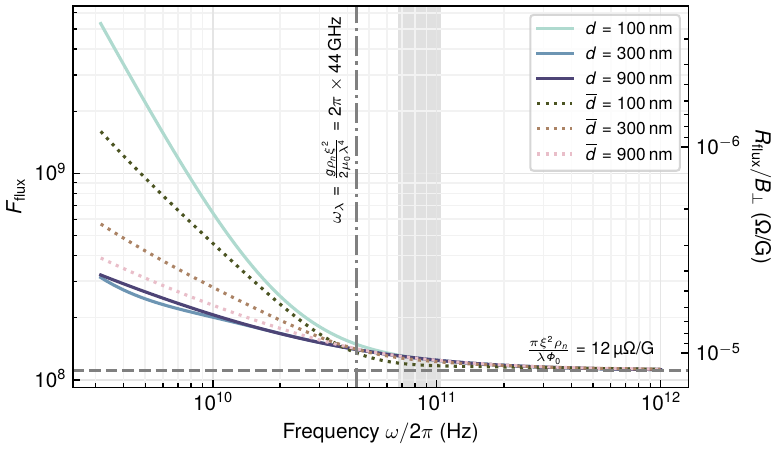}
    \caption[]{
    Comparison of trapped vortex loss
    as a function of drive frequency
    for various distributions of
    pinning impurity depth~$d$.
    We show vortex loss
    as a limiting finesse~$F_\text{flux}$
    at the measured magnetic field~$B_\perp = \SI{0.22}{\gauss}$
    and as surface resistance per unit field~$R_\text{flux}/B_\perp$.
    Solid lines consider impurities all at a fixed depth~$d$,
    while dotted lines take $d$
    to be drawn from an exponential distribution with mean~$\mean{d}$,
    as would result from impurities distributed as a Poisson point process.
    Though vortex loss is sensitive to impurity depth~$d$ at low frequencies,
    the dependence vanishes above a characteristic frequency~$\omega_\london$,
    including in the frequency range shown in Fig.~\ref{fig:cavity-perf}
    (gray shading).}
    \label{fig:vortex-loss}
\end{figure}

The resistance~$R_\text{flux}$ also varies
with the distribution of depths~$d$
of vortex-pinning impurities in the material
and with the drive frequency~$\omega$,
owing to the interplay of elastic and dissipative dynamics
in the motion of the flux tubes.
To estimate the finesse bound shown in Fig.~\ref{fig:cavity-perf}\ref{it:best-fins},
we perform a full numerical calculation~\cite{gurevich2017theory}
assuming that the depths~$d$
are exponentially distributed,
as would be true for impurities distributed as a Poisson point process,
with mean~$\mean{d}$
set by the mean free path~$\ell = \SI{320}{\nm}$~%
(Eq.~\ref{eq:drude}).

The precise choice of distribution is unimportant:
the cavity operates in a regime
with little dependence on impurity depth or drive frequency.
Indeed, calculating the full frequency-dependent surface resistance
for various impurity depth distributions
(Fig.~\ref{fig:vortex-loss})
we see that the residual surface resistance~$R_\text{flux}$
saturates at the high drive frequencies~$\omega$
used in this work
to a constant value
independent of impurity depth~$d$.
The limiting value
\begin{equation}
    \label{eq:vortex-highfreqlim}
    \frac{R_\text{flux}(\omega\to\infty)}{B_\perp}
    =
    \frac{\pi\xi^2\rho_n}{\london\Phi_0}
   \approx
   \SI{12}{\frac{\micro\ohm}{\gauss}}
\end{equation}
depends only on material properties
(coherence length~$\xi$,
penetration depth~$\london$,
and normal-state resistivity~$\rho_n$)
and the magnetic flux quantum~$\Phi_0 = h/2e$ (where $e$ is the elementary charge),
and is attained when the drive frequency~$\omega$
exceeds a characteristic frequency
\begin{equation}
    \omega_\london =
    \frac{g\rho_n \xi^2}{2\mu_0 \london^4}
    =
    2\pi\times \SI{44}{\GHz},
\end{equation}
where $g = \frac{1}{2} + \log \frac{\london}{\xi}$.



\section{Coherence of atom--light coupling}
\label{sec:atoms-numbers}

This section provides further details on the projected atom--cavity coupling and its implications for the achievable fidelity of nonlocal gates.  In Sec.~\ref{sec:geometric-eta},
we review the geometric expression for cooperativity~(Eq.~\ref{eq:eta-geom}) and its relation to the parameters~$(g, \kappa, \Gamma)$,
following Refs.~\cite{tanji2011interaction,haroche2006exploring}.  We proceed in Sec.~\ref{sec:cqed-params} to calculate the projected values of these parameters for our cavity when coupled to circular Rydberg states.  Finally, Sec.~\ref{sec:fidelity} explains the fundamental scaling of gate fidelity with cooperativity and analyzes an explicit scheme for implementing a cavity-mediated iSWAP gate, which reaches a 98\% fidelity for our parameters.

\subsection{Geometric expression for cooperativity}
\label{sec:geometric-eta}

In the main text, we give two expressions for the cooperativity~$\eta$.  The first,
\begin{equation}
    \label{eq:eta-def}
    \eta = \frac{4g^2}{\kappa\Gamma},
\end{equation}
defines the cooperativity in terms of ratios of the atom--photon coupling~$g$ to the cavity and atom decay rates $\kappa, \Gamma$.
We also rewrite the cooperativity in terms of geometric parameters for Fabry--P\'erot cavities (Eq.~\ref{eq:eta-geom}) to guide our cavity design.  In particular, for a cavity mode of finesse~$F$, waist~$w_0$, wavelength~$\lambda$, and Rayleigh range~$z_R = \pi w_0^2/\lambda$ coupled to a cycling transition in an atom at cavity center, the cooperativity is
\begin{equation}
    \label{eq:eta-geom-supp}
    \eta_0 = \frac{6}{\pi^3} \frac{F\lambda^2}{w_0^2}
    =
    \frac{6}{\pi^2} \frac{F\lambda}{z_R},
\end{equation}
assuming that the atomic transition and cavity mode have matching polarization.  Below, we derive this geometric expression from the definition of Eq.~\ref{eq:eta-def} and outline how it may be generalized beyond the ideal case of a cycling transition or extended to account for polarization mismatch.

First, we observe that the geometric expression in Eq.~\ref{eq:eta-geom-supp} does not depend on any parameters specific to the atomic transition.  We understand this result intuitively by interpreting the cooperativity as a ratio $\eta = \Gamma_c/\Gamma$ of the emission rate $\Gamma_c$ of an excited atom into the cavity to the spontaneous emission rate~$\Gamma$ of the same atom in free space.  Evaluating~$\Gamma_c = 4g^2/\kappa$ by Fermi's Golden Rule yields precisely the definition in Eq.~\ref{eq:eta-def}.  For a two-level atom, the two rates $\Gamma_c$ and $\Gamma$ have an identical dependence~$\Gamma_{(c)}\propto \abs{\mb d}^2$ on the dipole matrix element~$\mb d$ of the transition. The resulting maximal cooperativity~$\eta_0$ is therefore an intrinsically geometric quantity independent of~$\mb d$.


To derive the cooperativity~$\eta_0$ in Eq.~\ref{eq:eta-geom-supp}, we express the vacuum Rabi frequency~$g$ in terms of cavity parameters.  By definition,
\begin{equation}
    \label{eq:jaynes-cummings-g}
    g \equiv \frac{\Abs{\mb d^* \cdot \mb E_{\rms}}}{\hbar},
\end{equation}
where $\mb E_{\rms}$~is the rms vacuum field amplitude at cavity center. For a mode of frequency~$\omega$, this vacuum field is given by
\begin{equation}
    \label{eq:vacuum-field}
    \mb E_{\rms} = \sqrt{\frac{\hbar\omega}{2\epsilon_0 V_\text{mode}}}\bs{\epsilon}_c,
\end{equation}
where $\bs{\epsilon}_c$~is the field polarization and $V_\text{mode} = \frac{\pi}{4} w_0^2 L$~is the mode volume for the fundamental ($\tem_{00}$) Gaussian mode of a Fabry--P\'erot cavity of length~$L$.  We thus obtain the vacuum Rabi frequency
\begin{equation}
    \label{eq:geometric-g}
    g = \abs{\mb d}\sqrt{\frac{2\omega}{\pi\hbar\epsilon_0 w_0^2 L}}\abs{\bs{\epsilon}_a^* \cdot\bs{\epsilon}_c},
\end{equation}
where we have factored the dipole matrix element~$\mb d = \abs{\mb d} \bs{\epsilon}_a$ into a scalar $\abs{\mb d}$ and the polarization vector~$\bs{\epsilon}_a$ of the atomic transition.



Now consider the cavity emission rate~$\Gamma_c = 4g^2/\kappa$ and the atomic decay rate~$\Gamma$, which together will give the cooperativity~$\eta = \Gamma_c/\Gamma$.
We express the cavity linewidth~$\kappa$ in terms of finesse~$F$
using the free spectral range~$\omega_\text{FSR} = 2\pi c/2L$ of the cavity:
\begin{equation}\label{eq:kappaF}
    \kappa = \frac{\omega_\text{FSR}}{F} = \frac{\pi c}{L F}.
\end{equation}
Combining Eqs.~\ref{eq:geometric-g}--\ref{eq:kappaF} and assuming matched polarizations, such that $\abs{\bs{\epsilon}_a^* \cdot\bs{\epsilon}_c}=1$, we obtain
\begin{equation}
    \label{eq:cavity-scattering}
    \Gamma_c = \frac{4g^2}{\kappa} = \frac{8 \abs{\mb d}^2 \omega}{\epsilon_0 \hbar c} \frac{F}{\pi^2 w_0^2}.
\end{equation}
For a cycling transition, the atomic spontaneous emission rate~$\Gamma$ depends on the same dipole matrix element~$\mb d$ as 
\begin{equation}
    \label{eq:gamma}
    \Gamma = \frac{\abs{\mb d}^2 \omega^3}{3\pi\epsilon_0 \hbar c^3}.
\end{equation}
We thus arrive at the maximal cooperativity
\begin{equation}
    \eta_0 = \frac{\Gamma_c}{\Gamma} = \frac{6}{\pi^3} \frac{F\lambda^2}{w_0^2}.
\end{equation}

More generally, if the polarizations of the atomic transition and the cavity field are mismatched, the cooperativity is reduced to a value $\eta = \eta_0 \abs{\bs{\epsilon}_a^* \cdot \bs{\epsilon}_c}^2$.   Mismatched polarizations naturally arise when working with circular Rydberg state transitions, since the same dipole selection rules conferring these transitions their unity branching ratio also give them circular polarization, whereas the eigenmodes in high-finesse superconducting cavities tend to be linearly polarized.  In particular, the degeneracy of orthogonal linearly polarized modes is easily lifted by unintended astigmatism in cavities operating beyond the paraxial regime (Eq.~\ref{eq:post-parax-00}).  In our cavity, following the example of Ref.~\cite{kuhr2007ultrahigh}, we deliberately introduce a controlled splitting between linearly polarized modes by using toroidal mirrors.



The expression for the cooperativity can likewise be generalized beyond cycling transitions.  At mm-wave frequencies, this consideration is relevant for quantifying the cooperativity for low-angular-momentum states which, in contrast to circular states, are conveniently accessible via direct optical transitions from the ground state.  If the transition~$\ket{g} \leftrightarrow\ket{e}$ that couples to the cavity is not the only decay channel available to excited state~$\ket{e}$, then the cooperativity is reduced from its maximal value by the corresponding branching fraction: $\eta = \eta_0 (\Gamma_{eg} / \Gamma_e)$, where $\Gamma_{eg}$ is the decay rate on the $\ket{e} \rightarrow\ket{g}$ transition and $\Gamma_e$ the total decay rate of the excited state.


\subsection{Projected cavity QED parameters}
\label{sec:cqed-params}

\subsubsection{Circular Rydberg states}

To calculate the cooperativity of our cavity for coupling to circular Rydberg states, we use Eq.~\ref{eq:eta-geom-supp} and account for polarization mismatch as described above:
\begin{equation}
    \label{eq:eta-geom-pol}
    \eta = \eta_0 \abs{\bs{\epsilon}_a^* \cdot \bs{\epsilon}_c}^2 = \frac{6}{\pi^2} \frac{F\lambda}{z_R} \abs{\bs{\epsilon}_a^* \cdot \bs{\epsilon}_c}^2.
\end{equation}
To determine the wavelength~$\lambda$ for a transition between circular Rydberg states $\ket{nC}\leftrightarrow\ket{(n-1)C}$ in any hydrogenic atom, we use the Rydberg formula for the transition frequency $\omega = 2\pi c R\left[(n-1)^{-2} - n^{-2}\right]$, where $R$~is the Rydberg constant.  Choosing upper principal quantum number~$n = 42$ yields a frequency~$\omega = 2\pi\times \SI{92.08}{\GHz}$, sufficiently high to reach the finesse plateau~$F = \maxfinesse{}$ measured in the \lesscfcl{} and \hadriana{} cavity geometries.  Correspondingly, the transition wavelength is~$\lambda = 2\pi c/\omega = \SI{3.26}{\mm}$.  The cooperativity also depends on the Rayleigh range~$z_R$, which is a function of the cavity length~$L$ and the geometric parameter~$\mean{g}$~\cite{siegman1986lasers}:
\begin{equation}
    z_R = \frac{L}{2} \sqrt{ \frac{1+\mean{g}}{1 - \mean{g}} }.
\end{equation}
For the \hadriana{}-geometry, $z_R = \SI{21.1}{\mm}$. Finally, the overlap of the $\sigma^+$~atomic transition with a linearly polarized cavity mode yields $\abs{\bs{\epsilon}_a^* \cdot \bs{\epsilon}_c}^2 = \frac12$.  Therefore, we arrive at the cooperativity~$\eta = \num{2.72(4)e+6}$ given in the main text.



We additionally calculate the individual values of the coupling and decay parameters~$(g, \kappa, \Gamma)$, using the expressions given in Sec.~\ref{sec:geometric-eta}.  First consider the vacuum Rabi frequency~$g$ (Eq.~\ref{eq:jaynes-cummings-g}), which depends on the dipole matrix element and the vacuum field in the cavity.  For circular Rydberg state transitions~$\ket{(n-1)C}\leftrightarrow \ket{nC}$, the dipole matrix element scales in the large-$n$ limit as
\begin{equation}
    \mathbf{d} \approx \frac{1}{\sqrt{2}} n^2 e a_0 \boldsymbol{\epsilon}_+,
\end{equation}
where $e$ is the elementary charge,
$a_0$ the Bohr radius,
and the unit vector~$\boldsymbol{\epsilon}_+ = \frac{1}{\sqrt{2}}(1, i, 0)^T$
encodes the $\sigma^+$ polarization of the transition.
While this asymptotic expression gives $\mb d = 1247\, ea_0 \boldsymbol{\epsilon}_+$ for the transition of interest, we obtain a slightly smaller value~$1195\,ea_0\boldsymbol{\epsilon}_+$ from a numerical calculation~\cite{arc}.  The vacuum field is given by Eq.~\ref{eq:vacuum-field}, where the mode volume~$V_\text{mode} = \SI{0.81}{\cm^3}$ of the \hadriana{}-geometry yields a vacuum field amplitude~$\abs{\mb E_\rms} = \SI{2.1}{\mV/\m}$.  Multiplying the numerically calculated dipole matrix element~$\abs{\mb d}$ by the vacuum field amplitude~$\abs{\mb E_\rms}$, and accounting for the polarization overlap $\abs{\boldsymbol{\epsilon}_+\cdot\boldsymbol{\epsilon}_c} = \frac{1}{\sqrt{2}}$, we obtain
\begin{equation}
    g \equiv \frac{\abs{\mb d^* \cdot \mb E_\rms}}{\hbar} = 
    2\pi\times \SI{22.3}{\kHz}.
\end{equation}

The decay rates~$\kappa$ and $\Gamma$ are more straightforwardly found.
Ringdown measurements
directly give a cavity linewidth
\begin{equation}
    \kappa = 2\pi\times\SI{54.7(8)}{\Hz}
\end{equation}
in the \hadriana{}-geometry,
corresponding to the observed cavity finesse maximum~%
$\Fmax = \maxfinesse{}$.
The atomic linewidth~$\Gamma$
may be determined from Eq.~\ref{eq:gamma}
using the dipole matrix element~$\mb d$
calculated numerically above:
\begin{equation}
    \Gamma = 2\pi \times \SI{13.3}{\Hz}.
\end{equation}
The values $(g, \kappa, \Gamma) = 2\pi\times \SIlist{22e+3;55;13}{\Hz}$ reproduce the cooperativity calculated from the geometric expression of Eq.~\ref{eq:eta-geom-pol}, as expected.

\subsubsection{Optically accessible Rydberg states}

The cavity may alternatively be coupled to a transition between low-angular-momentum Rydberg states that are directly accessible via optical excitation from the ground state, at the expense of a reduction in cooperativity.  For example, in cesium, the transition between states
$\ket{g} \equiv \ket{36P_{3/2}, m_j = \tfrac{1}{2}}$
and
$\ket{e} \equiv \ket{37S_{1/2}, m_j = \tfrac{1}{2}}$
is at a frequency~\SI{95.974}{\GHz},
approximately the same as the circular-state transition chosen above to benefit from the maximum cavity finesse.
However, the dominant decay channel is now an optical transition to the ground state,
rather than the mm-wave transition that couples to the cavity.
Accounting for the branching ratio~$\Gamma_{eg} / \Gamma_e = \num{8.4e-4}$
yields a cooperativity~$\eta = \num{4.39(7)e+3}$,
with~$(g, \kappa, \Gamma) = 2\pi\times\SIlist{15e+3;55;3.7e+3}{\Hz}$~\cite{arc}.


\subsection{Forecasted gate fidelity}
\label{sec:fidelity}

One motivation for seeking a high cooperativity is to enable deterministic long-range entangling gates.   The cooperativity fundamentally limits the fidelity~$\Fid$ of a deterministic two-qubit gate to a value scaling as~$1-\Fid \propto 1/\sqrt{\eta}$~\cite{sorensen2003measurement,jandura2024nonlocal}.  In this section, we first provide an intuition for this limit by deriving it for the specific case of an iSWAP gate~\cite{schuch2003natural,mckay2016universal} generated by cavity-mediated spin exchange~\cite{majer2007coupling,norcia2018cavity,davis2019photon}.  Noting that past analyses of the scaling of gate fidelity with cooperativity in atomic systems~\cite{sorensen2003measurement,jandura2024nonlocal} have focused on level schemes relevant to optical cavities, we comment on subtleties in generalizing to transitions between Rydberg states.  We proceed to calculate the gate fidelity attainable in an explicit implementation of an iSWAP gate between circular Rydberg atoms in our cavity based on the parameters~$(g, \kappa, \Gamma)$.

\subsubsection{Cooperativity limit on gate fidelity}

To illustrate the limit set by the cooperativity on the fidelity of an entangling gate, we consider the example of a cavity-mediated spin-exchange process constituting an iSWAP gate.  In particular, we consider two atoms with qubit states~$\ket{\uparrow}, \ket{\downarrow}$ and transition frequency~$\omega_a$ coupled with vacuum Rabi frequency~$g$ to a cavity mode of frequency~$\omega_a = \omega_c + \Delta$.  The system is described by the Tavis--Cummings Hamiltonian
\begin{equation}\label{eq:Htc}
    \Htc = \frac{\Delta}{2}\sum_j \sigma_j^z + g\sum_j (\sigma_j^-a^\dagger + \sigma_j^+a),
\end{equation}
with $j$ indexing the qubits.  In the dispersive limit~$\Delta \gg g$, the cavity off-resonantly mediates spin-exchange interactions with coupling~$J = 2g^2/\Delta$.  Thus, after a time~$\Tgate = \pi / J = \frac{\pi\Delta}{2g^2}$ the pair states $\ket{\uparrow\downarrow}$~and~$\ket{\downarrow\uparrow}$ are exchanged and the cavity photon occupation remains unchanged, realizing an iSWAP gate~\cite{schuch2003natural,mckay2016universal}. While the conceptually simplest implementation operates with the cavity initialized in the vacuum state, the spin-exchange rate is independent of the intracavity photon number, providing robustness to thermal occupation of the cavity mode~\cite{sorensen2000entanglement}.

The fidelity of the gate is fundamentally limited by two sources of error: atomic decay at rates~$\Gamma_{\uparrow,\downarrow}$ for the two qubit states; and photon decay at rate~$\kappa$ from the cavity mode that mediates the interaction.  The probability~$\epsilon = 1-\Fid$ that an error occurs during the gate time~$\Tgate$ is
\begin{equation}
\label{eq:iSWAP_err}
\varepsilon \approx \left[\left(\frac{g}{\Delta}\right)^2\kappa + 2\br{\Gamma}\right] \Tgate = \pi\left(\frac{\kappa}{2\Delta} + \frac{\br{\Gamma} \Delta}{g^2}\right).
\end{equation}
where $\br{\Gamma} = (\Gamma_\uparrow + \Gamma_\downarrow)/2$~is the average decay rate for the two atomic states.  The error is minimized at an optimal detuning
\begin{equation}\label{eq:DeltaOpt}
    \DeltaOpt = g\sqrt{\frac{\kappa}{2\br{\Gamma}}} = \frac{\kappa}{2}\sqrt{\frac{\br{\eta}}{2}},
\end{equation}
where we define~$\br{\eta} = 4g^2/(\kappa\br{\Gamma})$ in terms of the average decay rate~$\br{\Gamma}$.  Choosing this optimal detuning yields a gate error
\begin{equation}\label{eq:error_from_eta}
    \varepsilon = 2\pi\sqrt{\frac{2}{\br{\eta}}}.
\end{equation}

A caveat is that reaching the optimal detuning (Eq.~\ref{eq:DeltaOpt}) while remaining in the dispersive limit requires that the cavity linewidth be broader than the atomic transitions, $\kappa \gg \br{\Gamma}$.  In optical cavities, this condition can generically be achieved by using the cavity as one leg of a Raman transition between two stable ground states, which effectively allows for tuning the linewidth of the (virtual) state that couples to the cavity.  By contrast, our millimeter-wave cavity will couple to a transition between two Rydberg states, both of which necessarily have a finite lifetime.  For example, letting the two qubit states be the circular states $\ket{\uparrow} = \ket{42 C}$ and $\ket{\downarrow} = \ket{41 C}$ yields an average atomic linewidth~$\br{\Gamma} = 2\pi \times \SI{14.2}{\Hz}$, only a few times smaller than the cavity linewidth~$\kappa = 2\pi\times \SI{55}\Hz$.  The resulting optimal detuning $\DeltaOpt \approx 1.4 g$ is only on the borderline of the dispersive regime.

Nevertheless, Eq.~\ref{eq:error_from_eta} provides a first estimate of the gate fidelity attainable with our cavity. Substituting the cooperativity~$\br{\eta} = 2.5\times 10^6$ for the $\ket{42 C}\rightarrow \ket{41 C}$~transition yields a fidelity~$\Fid = 1-\varepsilon = 0.994$.  Achieving this fidelity in practice may require optimal time-dependent control of the detuning, e.g., via Stark shifts induced by microwave or DC electric fields, to implement a fast near-resonant gate while ensuring that the cavity is ultimately left unpopulated.  In Ref.~\cite{jandura2024nonlocal}, such optimal control has been applied to achieve gate errors scaling as~$\varepsilon \sim 1/\sqrt{\eta}$ over a wide range of cooperativities and ratios~$\kappa/\Gamma$ of cavity to atomic decay rates, for both dispersive and near-resonant schemes, albeit in a level scheme with one stable ground state.

\subsubsection{Numerical simulation of iSWAP gate}

To establish the fidelity attainable in an explicit gate scheme with our parameters, we directly simulate the full Tavis--Cummings dynamics of two atoms coupled to the cavity at a fixed detuning $\Delta$.  We account for decay by evolving under a non-Hermitian Hamiltonian
\begin{equation}
H_\mathrm{eff} =  \Htc - \frac{i}{2}\Bigl(\kappa a^\dagger a + \sum_{j,\mu} \Gamma_\mu \ketbra{\mu}{\mu}_j\Bigr),
\end{equation}
where $\Htc$ is the Tavis--Cummings Hamiltonian of Eq.~\ref{eq:Htc},
$\mu \in \{\uparrow, \downarrow \}$ denotes the qubit state,
and $j$ indexes the atoms.  The resulting propagator~$V_t = \exp(-i \Heff t)$ describes the dynamics conditioned on no decay, and its non-unitarity quantifies the reduction in the probability that no decay has occurred as a function of time $t$.

To compare the resulting evolution with the target iSWAP gate,
\begin{equation}
U = i \left(\ketbra{\downarrow\uparrow}{\uparrow\downarrow} + \ketbra{\uparrow\downarrow}{\downarrow\uparrow}\right),
\end{equation}
we must additionally perform a projection~$\Pi$ onto the space of two-qubit states with no photon in the cavity.  We calculate the resulting gate fidelity according to Ref.~\cite{pedersen2007fidelity}:
\begin{equation}
    \altmathcal{F} = \frac{1}{D(D + 1)}\left[\tr(M M^\dagger) + \abs{\tr M}^2\right]
\end{equation}
where $D = 4$ is the dimension of the two-qubit Hilbert space, $M = \Pi U_0^\dagger V_t \Pi$ with $U_0$~the preimage under projection~$\Pi$ of the perfect iSWAP gate~$U$, and where we additionally allow for an arbitrary global spin rotation about the $z$-axis.

\begin{figure}
    \centering
    \includegraphics[width=0.5\textwidth]{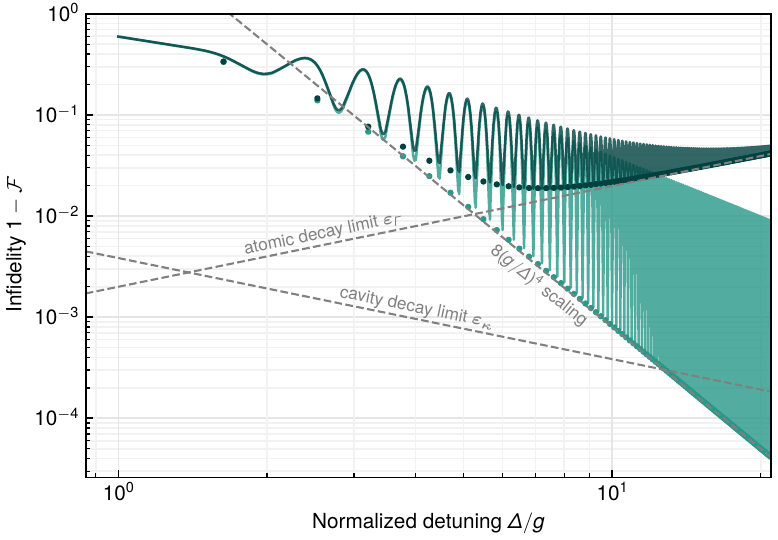}
    \caption{Infidelity~$1-\altmathcal{F}$ of cavity-mediated iSWAP gate versus normalized detuning~$\Delta/g$ for parameters~$(g, \kappa, \Gamma_\uparrow, \Gamma_\downarrow) = \SIlist{22e+3;55;13;15}{\Hz}$.  Continuous curves show evolution for time~$t = \tau_g = \pi \Delta/2 g^2$, while dots are for special detunings~$\Delta_m$
    evolving for times~$t_m$
    (Eq.~\ref{eq:special_vals}).
    Light green curves and markers are calculated for unitary evolution under~$\Htc$,
    while dark green curves and markers account for atomic and cavity decay by evolution under~$\Heff$.
    Also shown are the asymptotic scalings in the limit of large detuning~$\Delta \gg g$
    of error rates
    $\varepsilon_\TC \approx 8(g/\Delta)^4$ under coherent Tavis--Cummings evolution, $\varepsilon_\Gamma$ due to atomic decay,
    and $\varepsilon_\kappa$ due to cavity decay.}
    \label{fig:fid-estimate}
\end{figure}

The resulting infidelity is shown in Fig. \ref{fig:fid-estimate} as a function of detuning $\Delta$.  First, we fix the evolution time for each detuning to~$t = \tau_g = \pi/J$, where $J = 2 g^2/\Delta$~is the spin-exchange coupling in the dispersive limit.   We plot the resulting gate error for both the ideal unitary Tavis--Cummings dynamics (light green curve) and the dissipative evolution under~$H_\mathrm{eff}$ (dark green curve).  In both cases, the error exhibits oscillations as a function of detuning,  which are closely related to temporal oscillations in the cavity occupation at frequency~$\Delta$ for $\Delta \gg g$.  Local minima in the infidelity occur at values~$\Delta$ chosen so that the cavity is depopulated at time~$\tau_g$.

To more fully understand the dynamics, we note that the iSWAP gate can be equivalently viewed as a $\pi$~phase accruing between the triplet and singlet states $\ket{\pm} = (\ket{\uparrow\downarrow} \pm \ket{\downarrow\uparrow})/\sqrt{2}$.  In particular, a pair of atoms initialized in the triplet state with no photon in the cavity, $\ket{+}\otimes\ket{0}$, can emit a photon into the cavity to access the state~$\ket{\downarrow\downarrow}\otimes\ket{1}$, whereas the singlet state does not couple to the cavity.  On a Bloch sphere spanned by $\ket{+}\otimes\ket{0}$~and~$\ket{\downarrow\downarrow}\otimes\ket{1}$, the initial triplet state precesses at frequency~$\Omega = \sqrt{8g^2 + \Delta^2}$, acquiring a geometric phase~$\Phi = \pi(1-\Delta/\Omega)$ in each period of the precession.  Executing an iSWAP gate requires that the total evolution time~$t_m = 2\pi m/\Omega$ produce an integer number~$m$ of precessions and that the geometric phase accrued in each precession be~$\Phi = \pi/m$.  These conditions are both satisfied at detunings and corresponding evolution times 
\begin{subequations}
\label{eq:special_vals}
\begin{align}
\Delta_m &= g(m-1)\sqrt{\frac{8}{2m-1}}, \label{eq:special_dets}\\
t_m &= \frac{2\pi(m-1)}{\Delta_m}.\label{eq:special_times}
\end{align}
\end{subequations}

The gate errors for these detunings~$\Delta_m$, calculated at times~$t_m$, are shown by the circular markers in Fig.~\ref{fig:fid-estimate}.
The light green markers show the result for no dissipation,
while the dark green markers additionally account for atomic and cavity decay.
At small detuning, even the unitary evolution under~$\Htc$ produces an imperfect iSWAP operation, due to a nonzero probability that the cavity is occupied at time~$t_m$ if the qubits were initially in states~$\ket{\uparrow\uparrow}$.
The resulting error decreases with increasing detuning as~$\varepsilon_{\TC} \approx 8(g/\Delta)^4$ for $\Delta \gg g$.
At large detuning, corresponding to long evolution time~$t$, the dominant source of error becomes atomic decay with probability~$\varepsilon_\Gamma = 1- e^{-2\br{\Gamma}t}$.
The tradeoff between coherent errors~$\varepsilon_{\TC}$ and atomic decay~$\varepsilon_\Gamma$ results in an optimum detuning~$\Delta_m = 7.5 g$ (for $m = 15$).
Here, the infidelity reaches a minimum value~$1-\Fid = 1.9\times 10^{-2}$.
Only at even smaller detuning does cavity decay, with probability~$\varepsilon_\kappa \approx 1-e^{-(g/\Delta)^2\kappa t}$, become comparable to atomic decay.
Thus, the fidelity can likely be further be optimized by optimal control techniques that accelerate the gate by transiently populating the cavity~\cite{martinis2014fast,jandura2024nonlocal}.


\end{document}